\documentclass{aa}  

\usepackage{graphicx}
\usepackage{txfonts}

\usepackage[colorlinks=true,urlcolor=blue,citecolor=blue,linkcolor=blue]{hyperref}

\graphicspath{{./}{figures/}}

\begin{document}

   \title{The Cetus-Palca stream: A disrupted small dwarf galaxy}
   \subtitle{A prequel to the science possible with WEAVE with precise spectro-photometric distances}

   \author{Guillaume F. Thomas\inst{1,2}, Giuseppina Battaglia\inst{1,2}}
   \institute{Instituto de Astrofísica de Canarias, E-38205 La Laguna, Tenerife, Spain \and Universidad de La Laguna, Dpto. Astrofísica, E-38206 La Laguna, Tenerife, Spain\\
              \email{gthomas@iac.es}
             }

  \abstract
   {We present a new fully data-driven approach to derive spectro-photometric distances based on artificial neural networks. The method was developed and tested on Sloan Extension for Galactic Understanding and Exploration survey (SEGUE) data and will serve as a reference for the {\it Contributed Data Product} {\sc SPdist} of the William Hershel Telescope Enhanced Area Velocity Explorer (WEAVE) survey. With this method, the relative precision of the distances is of $\sim 13 \%$. The catalogue of more than 300 000 SEGUE stars for which we have derived spectro-photometric distances is publicly available on the {\it Vizier} service of the Centre de Donn\'ees de Strasbourg. With this 6D catalogue of stars with positions, distances, line-of-sight velocity, and {\it Gaia} proper motions, we were able to identify stars belonging to the Cetus stellar stream in the integrals of motion space. Guided by the properties we derived for the Cetus stream from this 6D sample, we searched for additional stars from the blue horizontal and red giant branches in a 5D sample. We find that the Cetus stream and the Palca overdensity are two parts of the same structure, which  we therefore propose to rename the Cetus-Palca stream. We find that the Cetus-Palca stream has a stellar mass of $\simeq 1.5 \times 10^6$ M$_\odot$ and presents a prominent distance gradient of 15 kpc over the $\sim 100 \degr$ that it covers on the sky. Additionally, we also report the discovery of a second structure almost parallel to the Cetus stream and covering $\sim 50\degr$ of the sky, which could potentially be a stellar stream formed by the tidal disruption of a globular cluster that was orbiting around the Cetus stream progenitor.}


   \maketitle

\section{Introduction}
Large astrometric and spectroscopic surveys, such as {\it Gaia} \citep{gaiacollaboration_2016,gaiacollaboration_2018}, the William Hershel Telescope Enhanced Area Velocity Explorer \citep[WEAVE;][]{dalton_2012}, 
the Dark Energy Spectroscopic Instrument \citep[DESI;][]{flaugher_2014}, Sloan Digital Sky Survey \citep[SDSS;][]{kollmeier_2017},  Large Sky Area Multi-Object Fibre Spectroscopic Telescope \citep[LAMOST;]{zhao_2012a}, or the
Hectochelle in the Halo at High Resolution \citep[H3;][]{conroy_2019a}, are providing chemical and kinematic properties for a tremendous number of individual stars in our Galaxy. With these data, it is possible to reconstruct the formation history of the Milky Way  in detail \citep[e.g.][]{belokurov_2018a,haywood_2018,kruijssen_2020,naidu_2020,fernandez-alvar_2021,ishigaki_2021,lane_2021,malhan_2021,simpson_2021} and to compare it to the formation histories of similar galaxies in large-scale cosmological simulations. Precise distances of individual stars are a crucial ingredient in the determination of the structure and kinematic properties of the Milky Way stellar component; for example to transform {\it Gaia} proper motions into physical tangential velocities. However, despite the great advances brought about by the European Space Agency’s {\it Gaia} mission to provide parallaxes for more than 1.4 billions stars \citep{gaiacollaboration_2016,gaiacollaboration_2020}, these are not of sufficient precision to  measure the distances of stars beyond $\sim10$ kpc \citep[e.g.][]{ibata_2017a}, such as those located in the outer disc or in most of the stellar halo;  other methods are therefore needed to enable an investigation of the properties of the outskirts of the Milky Way. 

Several methods have been developed to infer distances, either using only photometric data \citep[e.g.][]{juric_2008,ivezic_2008,deason_2011,ibata_2017a,sesar_2017,thomas_2019,conroy_2021} or a combination of spectroscopic and photometric data \citep[e.g.][]{xue_2014,coronado_2018,mcmillan_2018,queiroz_2018,hogg_2018,cargile_2020}. Some developed methods statistically infer the distances of the stars based on the assumptions made on the global distribution of the stars in the Galaxy \citep[e.g.][]{bailer-jones_2015,bailer-jones_2018,queiroz_2018,anders_2019,pieres_2019}. However, these methods depend sensitively on the adopted Galactic spatial distribution prior \citep{hogg_2018}, which is still not precisely known, especially in the stellar halo, as different tracers yield different trends \citep{thomas_2018a,fukushima_2019}. The methods to obtain photometric distances do not require (expensive) spectroscopic measurements, but are often limited to specific stellar populations. These are not always easy to identify photometrically, leading to potentially strong biases. Spectrophotometric methods tend to be more accurate because they use both photometric and spectroscopic information. However, most of them are still focused on specific stellar populations (K-giants: \citealp{xue_2014}, main sequence (MS) stars: \citealp{coronado_2018}, red giant branch (RGB) stars: \citealp{hogg_2018}), and the majority of them are based on theoretical spectro-photometric relations  \citep[e.g.][]{xue_2014,mcmillan_2018,queiroz_2018,cargile_2020}.

Thanks to the {\it Gaia} early third data release \citep[EDR3][]{gaiacollaboration_2020}, it is now possible to develop new methods to derive spectro-photometric distances using the large number of stars with spectroscopic data and precise parallaxes; this can potentially be done for stars from different stellar evolutionary phases and  without relying on theoretical spectro-photometric models.  Here, we
present a new data-driven method to precisely measure distances using spectroscopic and photometric information using machine learning (ML) techniques. This method is heavily based on the previous work described in \citet{thomas_2019}; the major difference is the use here of spectroscopic parameters (gravity, effective temperature and metallicity), which improve the distance measurements, especially for giant stars, and the absence of the $u$-band photometry from the Canada-France-Imaging-Survey \citep[CFIS/UNIONS][]{ibata_2017a}. Our primary objective was to develop the method and the algorithm that will be used to provide spectro-photometric distances for the stars observed by WEAVE as the WEAVE Contributed Data Product\footnote{A {\it Contributed Data Product} is a product beyond what the WEAVE Advanced Processing System will provide.} (CDP) {\sc SPdist}. However, at the time of writing, WEAVE is not yet operational. Therefore, we developed our method using stars observed by the Sloan Extension for Galactic Understanding and Exploration survey \citep[SEGUE][]{yanny_2009}, because among the publicly available spectroscopic datasets it is the closest to what WEAVE will deliver in terms of instrumental specifications in its low-resolution mode. The dataset, the architecture of the algorithm, and the validation of the precision on the distance achieved with this method are presented in Section~\ref{sec:spdist}. We applied this algorithm to derive spectro-photometric distances for more than 300 000 stars present in SEGUE, the catalogue of which is publicly available at Centre de Donn\'ees de Strasbourg (CDS). In Sect.~\ref{Sec_phase_space}, we present the identification of the Cetus stream as a structure in the integral-of-motion space using 6D information for SEGUE stars, including the spectro-photometric distances derived in this work. In Section~\ref{sec:cetus_erd3} we use the information gained by integrating the orbit of the stars of the Cetus stream as a guide to expand the search to the more numerous sample of stars with 5D information: first we focus on blue horizontal branch (BHB) stars (in Section~\ref{sec:BHB_cetus}), and then on the different stellar populations of Cetus present in {\it Gaia} EDR3. Finally, we present our conclusions in Section~\ref{sec:conclusion}.

\section{Determination of spectro-photometric distances} \label{sec:spdist}

In this section, we present a new ML-based method to determine the distance of individual stars through their spectro-photometric parameters. Contrary to \citet{cargile_2020}, who used mock data to fit their spectro-photometric distance relation, this method uses a data-driven approach.

\subsection{Data}

We train and apply the method to the stars observed by the SEGUE survey \citep{yanny_2009}. 
 From the SEGUE stellar catalogue, only stars with a photometric counterpart in the second data release of the Pan-STARRS1 3$\pi$ Steradian survey \citep[PS1][]{chambers_2016} and in {\it Gaia} EDR3 are used hereafter. More specifically, we used the forced-{\sc WARP} PSF photometry in the $griz$-bands from PS1 \citep{magnier_2020} and the $G$-band from EDR3. The PS1 $y$-band is not used because of its shallowness compared to the other bands, and {\it Gaia} $G_{BP}$ and $G_{RP}$ are not used because of the excess in colour at the faintest magnitudes correlated with large uncertainties. Although, this is not very important for SEGUE stars, because this survey has a depth of G$\sim 18$, it is important to take this point into account at an this stage because WEAVE is expected to go down to the {\it Gaia} magnitude limit in the Galactic archaeology subsurvey of the thick disc and stellar halo at low spectral resolution, as presented in the \href{https://www.ing.iac.es/confluence/download/attachments/1049035/WEAVE-SCI-002%20The%20WEAVE%20Science%20Case%20v3.2.pdf?version=1&modificationDate=1600251942118&api=v2}{WEAVE science case}.

All the stars are corrected for foreground extinction assuming the $E(B-V)$ values given by \citet{schlegel_1998} at their positions,  with the reddening conversion coefficients for the $griz$-bands given by \citet{schlafly_2011} for a reddening parameter of $R_v = 3.1$, while for the {\it Gaia} $G$ filter, we follow \citet{sestito_2019} by adopting the coefficients from \citet{marigo_2008}. It should be noted that some stars from this cross-matched catalogue are affected by crowding in dense regions (i.e. globular clusters) especially with the PS1 forced-{\sc WARP} photometry used here. Therefore, to prevent any biases generated by those stars, and to remove extended objects such as background galaxies, we keep only the stars with |{\sc stargal}|$< 3$. The {\sc stargal} parameter is the median (in sigma) of the difference between the Kron and the PSF photometry compared to their uncertainties added in quadrature for all single exposures of a single object \citep[see][]{magnier_2020,magnier_2020a}.

\begin{figure*}
\centering
  \includegraphics[angle=0, viewport= 0 0 1025 570, clip,width=14cm]{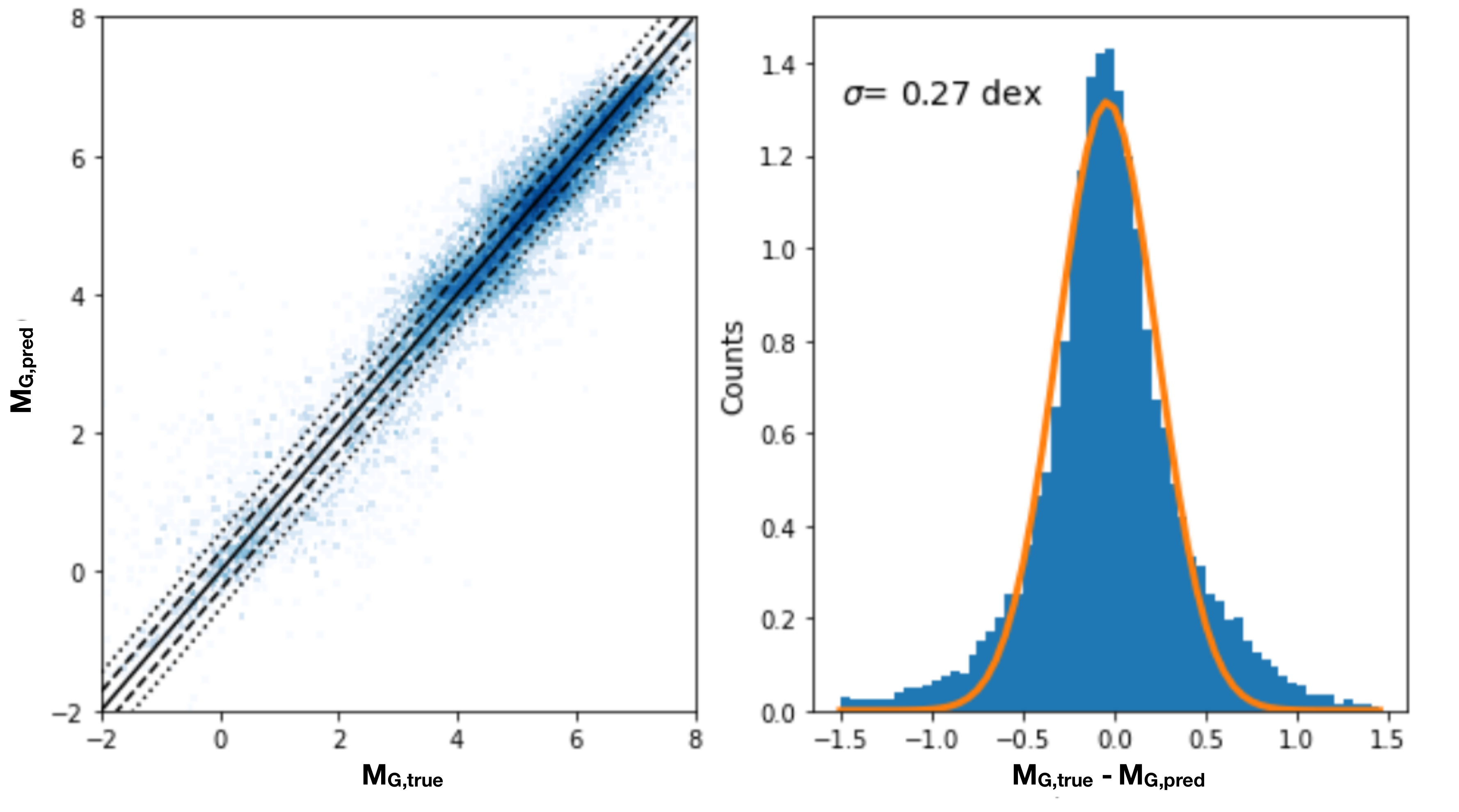}
   \caption{Comparison of the `true' (M$_{G,true}$) and predicted (M$_{G,pred}$) absolute magnitude for the stars of the training sample. The left panel shows the true and derived quantities plotted against each other. The one-to-one relation is shown as a solid line, and the dashed and dotted lines correspond  to the $1-\sigma$ and $2-\sigma$ deviations, respectively. The right panel shows the distribution of the difference between the true and derived absolute magnitude, with a Gaussian fit overlaid. A scatter of 0.27 magnitude corresponds to a relative uncertainty on the distance of 13\%.} 
\label{Residual}
\end{figure*}

\begin{figure*}
\centering
  \includegraphics[angle=0, viewport= 0 0 1025 560, clip,width=17cm]{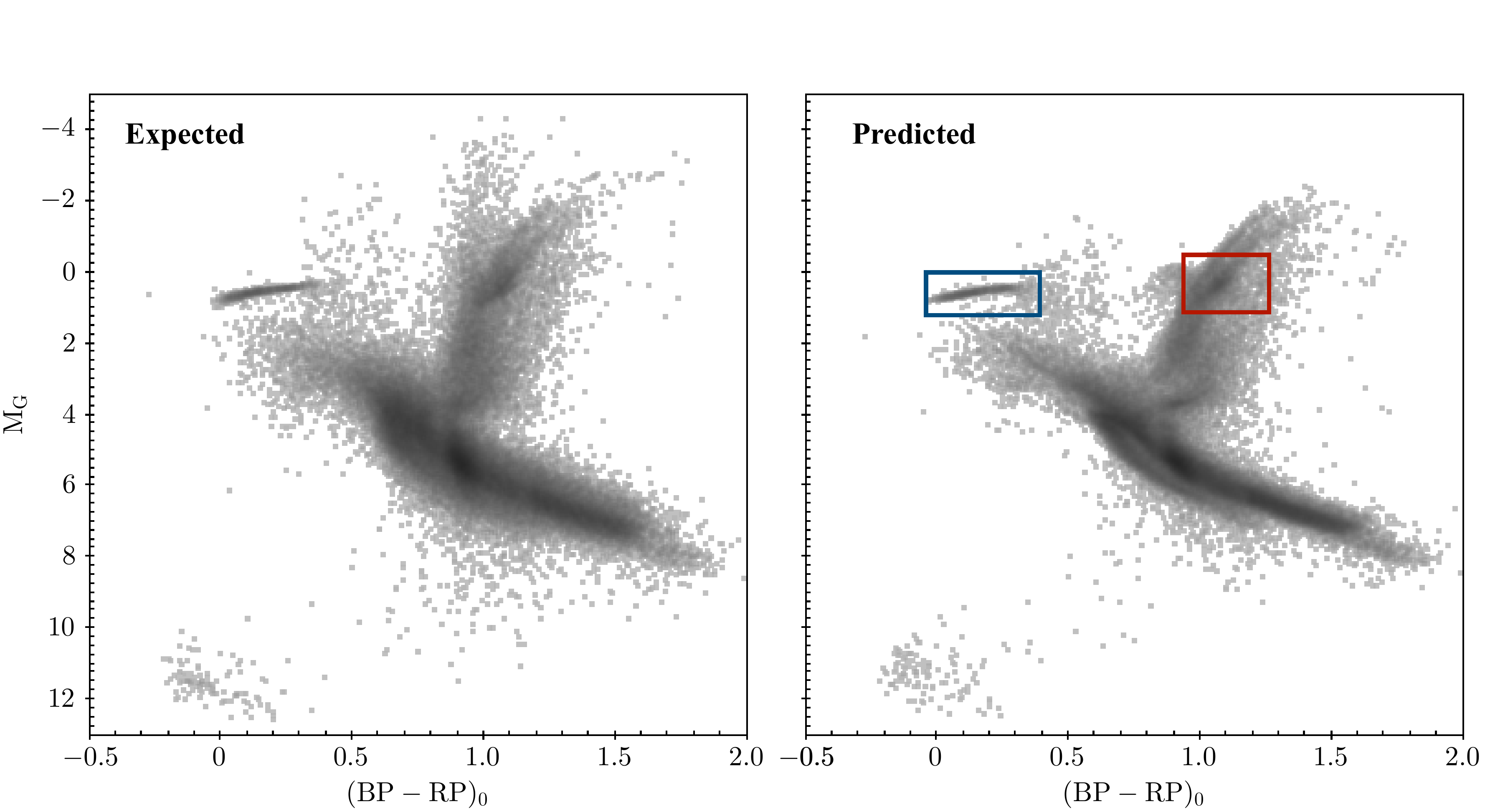}
   \caption{Colour--magnitude diagram of the training sample with the expected absolute magnitude in the {\it Gaia} $G$-band shown in the left panel and the absolute magnitude predicted by the algorithm in the right panel. It is striking to see the RC on the predicted CMD so clearly (highlighted by the red box), while this feature it is less obvious on the expected CMD. We also highlighted with the blue rectangle the location of the BHB stars, which were added in the training sample as explained in Section~\ref{sec:method}.} 
\label{CMD_training}
\end{figure*}

\subsection{The method} \label{sec:method}
 Our algorithm uses the spectroscopically determined effective temperature (T$_{eff}$), surface gravity ($\log(g)$), and metallicity ([Fe/H]) with a combination of photometric colours to estimate the absolute magnitude in the {\it Gaia} $G$-band for each star, and therefore its distance, through an artificial neural network (ANN). To perform this so-called regression, the inputs of the ANN are the {\sc teffadop}, {\sc loggadop,} and {\sc fehannrr} parameters from the SEGUE Stellar Parameters Pipeline \citep[SSPP][]{lee_2008}, and all the possible colours that can be produced from the $g_0$,$r_0$, $i_0$, $z_0$, and $G_0$ extinction-corrected photometry\footnote{i.e. $(g-r)_0$, $(g-i)_0$, $(g-z)_0$, $(r-i)_0$, $(r-z)_0$, $(i-z)_0$, $(g-G)_0$, $(G-r)_0$  $(G-i)_0$ and $(G-z)_0$}. We choose not to use the {\sc fehadop} metallicity because it is a combination of metallicity derived by different methods, which tend to erase the signal in the tails of the distribution \citep[see their Appendix A of ][]{starkenburg_2017}. Following \citet{starkenburg_2017}, we decided to use the {\sc fehannrr} parameter because it provides a more robust estimate, particularly in the low-metallicity regime, where the bulk of the halo stars are found. In order to improve the performance of the ANN and to limit the potential biases, all the input parameters are normalised to have a distribution with a mean equal to zero and a standard deviation of one with respect to the training sample.

To train and test an ANN, it is advantageous to use a training set that is as large as possible. However, the training set has to be composed of stars with precise parameters in order to prevent the algorithm from learning  a false relation   due to objects with imprecise parameters. Therefore, we imposed a signal-to-noise ratio (S/N) cut of S/N$\ge 20$ for the stars in the training sample. This threshold was chosen because at lower S/N, the distribution of the uncertainties on the parameters given by the SSPP as a function of the S/N is irregular, indicating that the parameters in that region are not reliable \citep[see discussion in ][]{thomas_2019}. 

Because the surface gravity plays a key role in disentangling  MS and RGB stars, which at the same colour have very different absolute magnitudes, we only consider stars with uncertainties $\delta \log(g)< 0.2$ dex for the training sample, which is just below the typical internal uncertainty on {\sc loggadop} \citep[0.19 dex according to][]{lee_2008}.

Furthermore, we kept only stars with photometric uncertainties in all the bands better than 0.1 mag.

The absolute magnitudes on which the ANN is trained are obtained from {\it Gaia} parallaxes ($\varpi$), such as $M_{G}= G_0 + 5 + 5\log_{10}(\varpi$ [mas]$/1000)$. We correct {\it Gaia} parallaxes from the zero-point off-set as a function of colour and sky location, following \citet {lindegren_2020}. This method imposes a positive zero-point-corrected parallax on the stars of the training sample. Thus, we only keep stars with a minimum parallax of $\varpi=0.1$ mas yr$^{-1}$, as stars with a smaller parallax tend to have unrealistic absolute magnitudes \citep{luri_2018,thomas_2019}.

As show by \citet{luri_2018}, the inversion of the parallax to obtain the distance (and so the absolute magnitude), is only valid for relative parallax uncertainties $\varpi/\delta \varpi \ge 5$ (or a relative distance precision of better than 20\%). For the dwarf stars ($\log(g) \ge 3.5$), which are largely from the  MS, this is not a problem because 104 516 dwarf stars respect this criterion, covering a wide range of metallicity and temperature. However, as mentioned by \citet{thomas_2019}, using this relative parallax precision criterion also on giant stars ($\log(g) < 3.5$) would limit the sample to only about $3000$ stars. Furthermore, these are not even representative of the full giant sample of the SEGUE dataset, as most are subgiant stars with $\log(g)\sim 3$. The reason for this difference between dwarf and giant stars is that at a similar apparent magnitude, the giants are more distant than the dwarfs, which leads to a higher uncertainty on their parallax. Thus, following \citet{hogg_2018} and \citet{thomas_2019}, instead of imposing a relative precision cut on the parallax for giant stars, we keep only those with $\delta \varpi < 0.07$ mas.

We also used $4625$ of the $6036$ K-giants from the catalogue of \citet{xue_2014} (the other $1400$ do not respect the photometric uncertainty criteria in all bands), which have distance measurements with a relative precision of 16\%. This additional dataset marks a great improvement compared to \citet{thomas_2019}, especially for the intrinsically brightest stars in SEGUE, by adding a large number of giant stars with precise distances. For the K-giants from the \citet{xue_2014} catalogue that pass the  $\delta \varpi < 0.07$ mas cut, we used the (maximum likelihood) distance modulus (DM) provided by \citet{xue_2014} rather than using the {\it Gaia} parallax to infer the absolute magnitude used to train the ANN. A total of 14 826 giant stars compose the training set, of which 10,201 have been selected based on their {\it Gaia} parallax.

The BHB stars present in the training sample, which can be identified through their location on the colour--magnitude diagram (CMD) in Fig.~\ref{CMD_training}, have large uncertainties on their parallax, leading to a broad range of `true' absolute magnitudes. We were surprised to find that our algorithm is able to recover a BHB, likely thanks to the input spectroscopic surface gravity. However, the intrinsic brightness tends to be underestimated, such that the distance of these BHBs is overestimated by a factor of $\sim 1.1$ when compared to the distances obtained from the relation of \citet{deason_2011} based on SDSS photometry in the $g$ and $r$-bands. We therefore decided to add BHB stars from the catalogue of \citet{xue_2008} to the training sample, with the absolute magnitude derived using the relation of \citet{deason_2011}. 
This finally leads to a training sample composed of 14 826 giant stars, 104 321 dwarf stars, and 1 376 BHB stars. The impact of adding the K-giants of \citet{xue_2014} and the BHBs of \citet{xue_2008} to the training sample on the precisions of the predicted distances is discussed in Appendix~\ref{app_A}.

By itself, the architecture of the ANN is rather classical. It is composed of four hidden layers composed of respectively 2048, 512, 64, and 32 neurons using a rectified linear unit (ReLU) activation function built with the {\sc Keras} package \citep{chollet_2015}. Due to the large number of possible outliers (especially for giant stars), the cost function used by the ANN is a mean squared-error (MSE) coupled to an adaptive moment estimation (also known as an Adam) optimisation method \citep{kingma_2014} to prevent falling to a local minimum.

The difference between the predicted and the `true' absolute magnitude (used to train the ANN) is shown in Figure \ref{Residual}. The residual is of $\sigma = 0.27$ mag, corresponding to a relative distance precision of 13\%, and does not show any trend with absolute magnitude. Notwithstanding the inhomogeneities of the training sample, it seems reasonable to think that the 13\% precision is an upper limit when comparing the CMD built from the  `true' absolute magnitude (Fig.~\ref{CMD_training},left) and that made from the predicted absolute magnitude (Fig.~\ref{CMD_training}, right). Indeed, not only can our algorithm recover and enhance the standard distance candles that are the BHB stars (within the blue rectangle) and the red clump (within the red rectangle), but even the MS is better characterised, with two almost parallel  well-defined branches corresponding to the Galactic disc and to the Galactic halo/Gaia-Enceladus-Sausage \citep{haywood_2018}. The only population for which our algorithm fails to improve the absolute magnitude prediction are the white dwarfs. This is due to an incorrect estimation of the surface gravity for the white dwarfs by the SSPP, with $\log(g)\simeq 5$ cm s$^{-2}$ instead of $\log(g) = 7-9$ cm s$^{-2}$ \citep[e.g.][]{tremblay_2013}. The precision of the spectro-photometric distances as a function of the input  spectroscopic parameters is discussed in Appendix~\ref{app_B}.

\begin{figure*}
\centering
  \includegraphics[angle=0, clip,width=12.5cm]{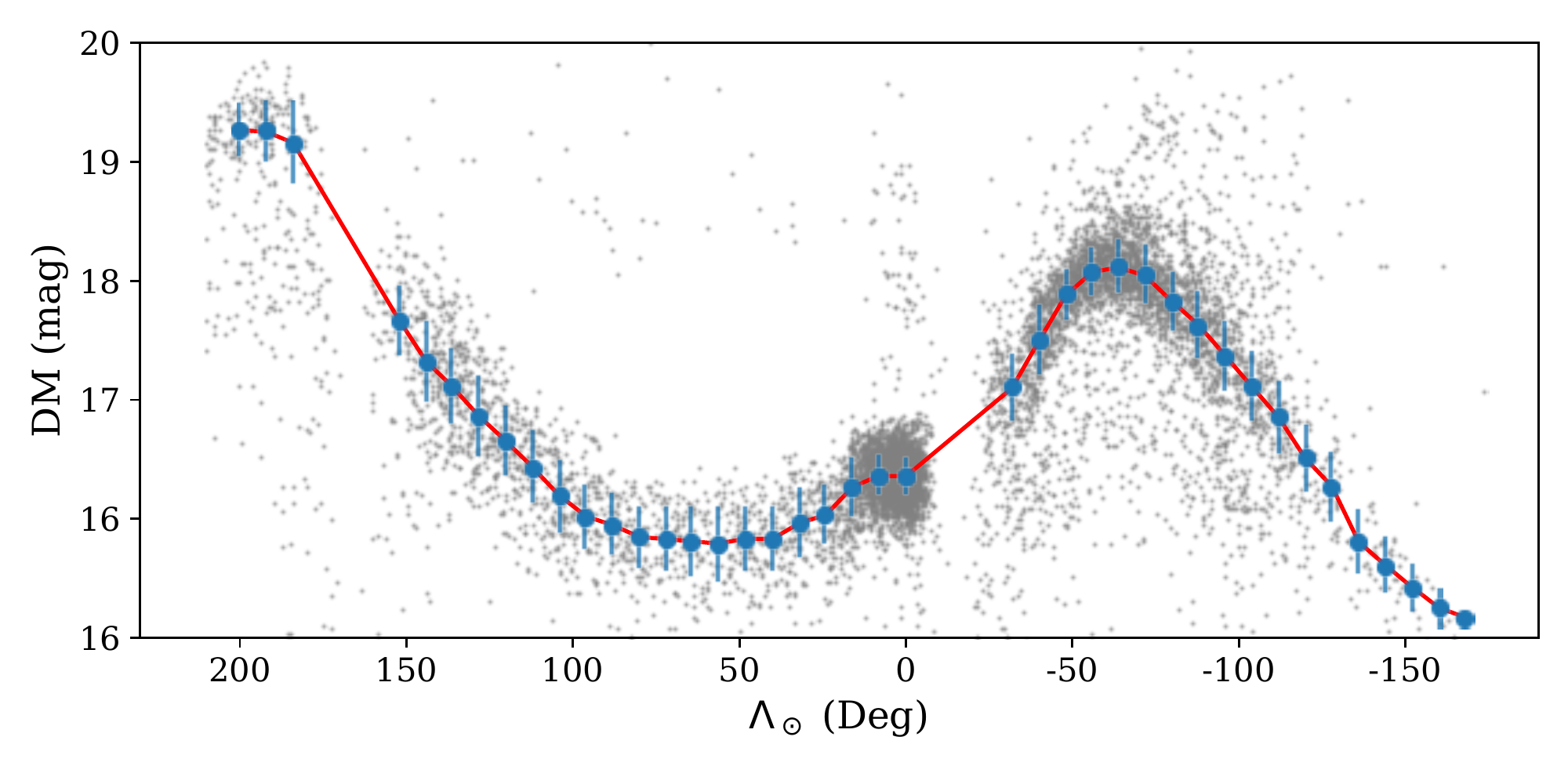}
    \includegraphics[angle=0, clip,width=12.5cm]{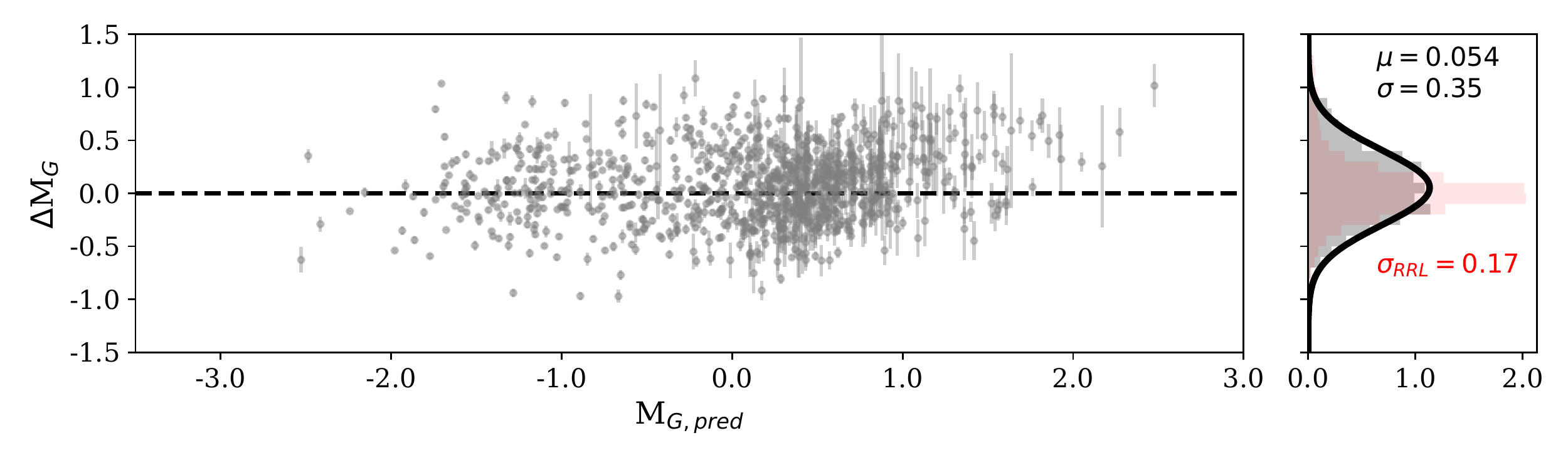}
   \caption{{\it Top panel:} Distance modulus of the Sgr RR Lyrae as a function of longitude along the plane of the Sagittarius stream, as defined by \citet{majewski_2003} (in grey). The blue error bars show the mean and standard deviation of the RR Lyrae DM as a function of the $\Lambda_\odot$ position, and are used to define the Sgr DM track represented by the red cubic spline function. {\it Bottom panels}: Difference between the absolute magnitude derived by the DM ridgeline and the absolute magnitude predicted by our algorithm as a function of the predicted absolute magnitude for stars along the Sgr stream (left). On the right panel, the red histogram shows the normalised residuals of the DM of the RR Lyrae to the fitted ridgeline along the Sgr stream.} 
\label{RRLdist}
\end{figure*}

Once the ANN was trained, we applied it to all stars from the initial cross-match SEGUE-PS1-EDR3 catalogue. The catalogue, composed of 308 692 stars with 13\% precise distance, is available at the CDS.

\begin{figure}
\centering
  \includegraphics[angle=0, clip,width=8.5cm]{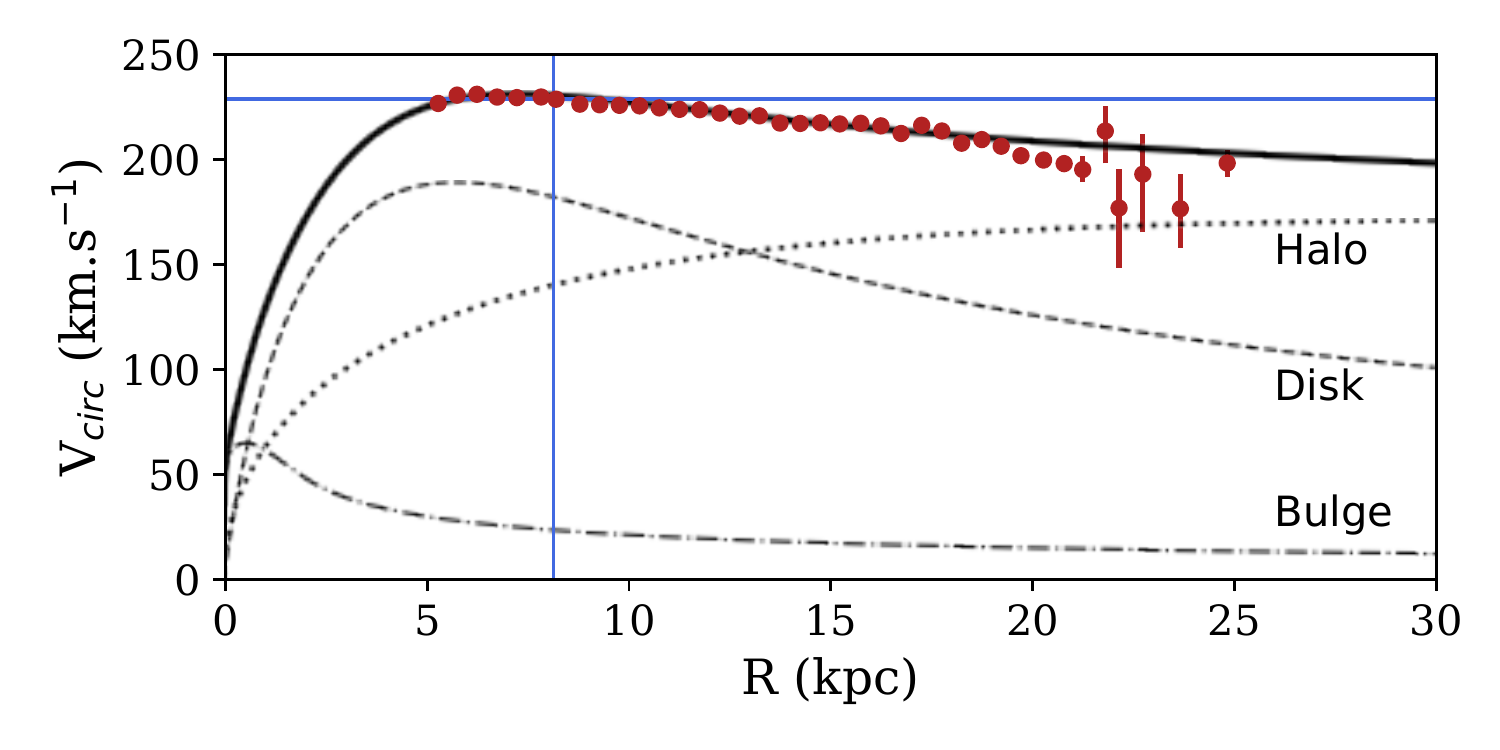}
   \caption{Rotation curve of the Milky Way model used in this work (thick line) together with the measured rotation curve of our Galaxy by \citet{eilers_2019} (red dots). The thin blue lines show the rotation velocity at the solar radius (229 km.s$^{-1}$ at 8.129 kpc).} 
\label{rotcurv}
\end{figure}

\begin{figure*}
\centering
  \includegraphics[angle=0, clip,width=15cm]{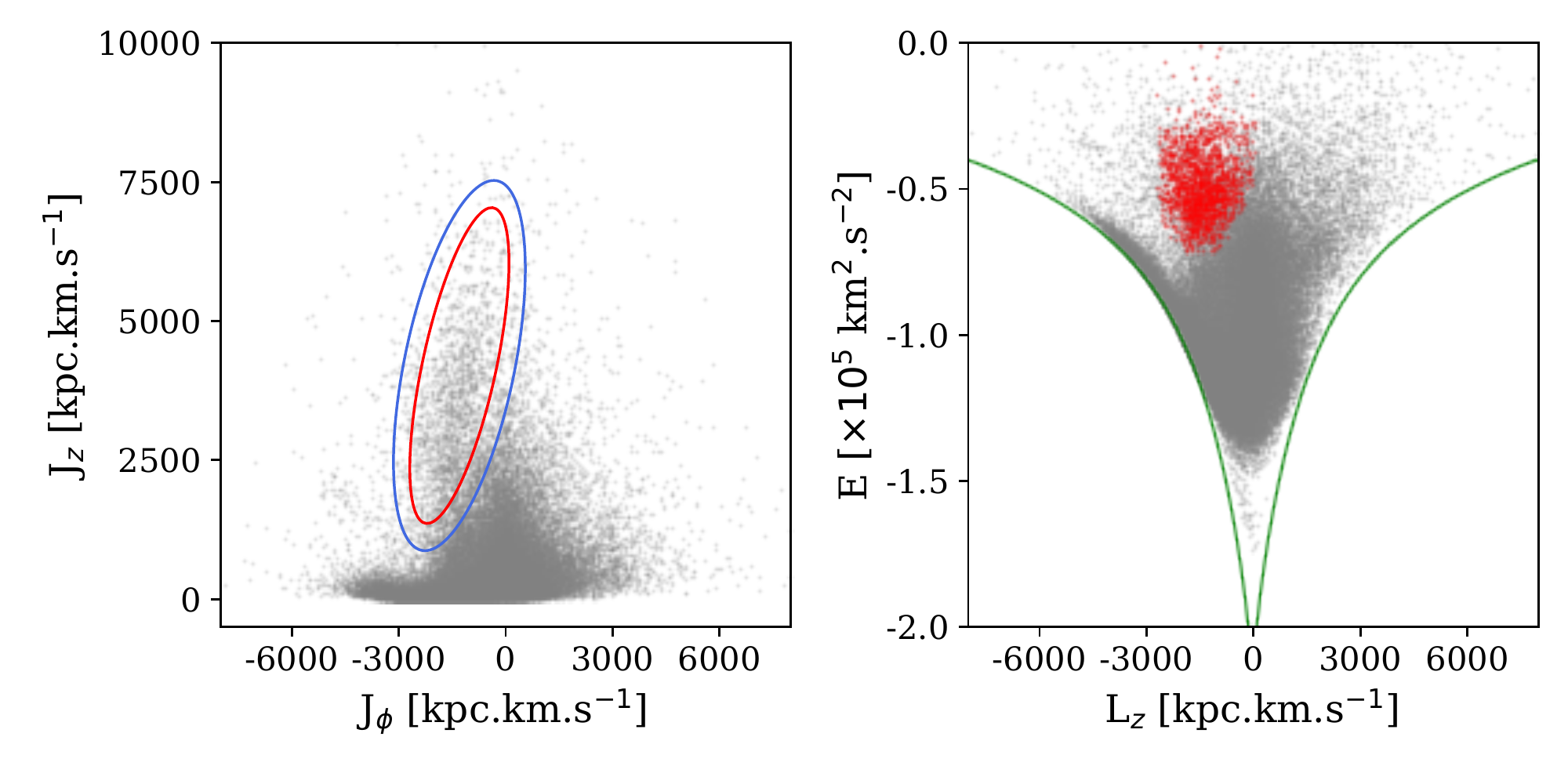}
   \caption{Azimuthal and vertical actions of the stars from the SEGUE dataset (left), and corresponding energy-vertical angular momentum diagram (right). The green lines mark the locus of circular orbits in the Milky Way model. The red ellipse on the left panel shows clearly visible structure corresponding to the Sagittarius stream. The stars inside that selection are highlighted in red in the right panel. The stars between the red and blue ellipses in the left panel are used to estimate the background contamination, which is used in Fig.~\ref{Sgr_pos}.} 
\label{phase-space} 

\end{figure*}

\subsection{Verification of the distance precision}

Though the distances of MS stars are, by construction of our training sample, well established, this is not necessarily the case for the giants, where a certain amount of doubt can persist regarding their precision. Here it is not possible to use globular clusters to obtain independent verification of the prediction of our algorithm because of the low number of globular cluster stars observed by SEGUE whose photometry is not significantly affected by crowding. However, a significant number of SEGUE stars are part of the Sagittarius (Sgr) stream, which covers a broad range of distances \citep[e.g.][]{koposov_2012,belokurov_2014}; we can therefore use this structure to verify the precision of our distances. 

We identify star members of the Sgr stream through their phase-space position (see Section \ref{Sec_phase_space}). The residual contaminants are removed using the selection criteria based on proper motions and position on the sky described in \citet{ibata_2020}. This leads to 888 SEGUE stars identified as part of the Sgr stream, mainly those located between $0 \degr<$ R.A. $<50\degr$ and $100 \degr<$ R.A. $<200\degr$, all of them being giant stars ($\log(g)$<3.5) either on the BHB or on the RGB.

To determine the DM variation along the Sgr stream, we use RRLyrae stars, because of their small typical uncertainties on the distance, which are estimated to be 3\% \citep{hernitschek_2018}. We applied the selection criteria of \citet{ibata_2020} in terms of position on the sky and proper motions to a catalogue of candidate RR Lyrae stars built as the union of the {\it Gaia} DR2 SOS gaiadr2.vari\_rrlyrae \citep{holl_2018, collaboration_2019,clementini_2019}, the stars classified as RRLyrae of a, b, c, or d type in the general variability catalogues gaiadr2.vari\_classifier\_result, and the PS1 RRLyrae of ab or cd type by \citet{sesar_2017}\footnote{With classification score above 0.6 and with a detection in {\it Gaia}.}. This leaves us with 10 384 potential Sagittarius stream RR Lyrae. The DM is determined assuming an absolute magnitude of $M_G=0.69$ mag calculated from the $M_G -$ [Fe/H] relation of \citet{muraveva_2018} for a mean metallicity of [Fe/H]$=-1.3$ \citep{ibata_2020}. The variation of the Sgr DM as a  function of the position along the stream was determined by fitting a cubic spline to the RR Lyrae track (Fig.~\ref{RRLdist}, top), where $\Lambda_\odot$ is the longitude along the Sagittarius stream orbital plane \citep{majewski_2003}.

This DM track was then used to determine the expected absolute magnitude of the 888 SEGUE stars along the Sgr stream. The difference between this and the absolute magnitudes predicted by our method is shown in the lower panel of Fig.~\ref{RRLdist}. The standard deviation is of 0.35 mag, which is larger than the 0.27 mag found previously. However, this also includes the intrinsic width of the Sgr stream, as well as the scatter introduced by variations in the RRLyrae metallicity and the M$_G$-[Fe/H] calibration itself. The dispersion of the Sagittarius stream RRL absolute magnitudes  around the DM track amounts to 0.17 mag (red histogram in Fig.~\ref{RRLdist}). By deconvolving the dispersion of the absolute magnitudes of SEGUE giant members of the Sagittarius stream with this value, we obtain 0.31 mag, leading to uncertainties on the distance of 14\%, with a negligible bias of 0.054 mag; this confirms the good accuracy of our algorithm. A 14\% relative uncertainty on distance is probably an upper limit, as one can appreciate a small amount of contamination even after the position and proper motion cleaning, where this would lead to a larger spread around the DM track. For the remainder of the paper, we continue assuming typical uncertainties of 13\%.

We note that if we were to use an absolute magnitude of $M_G=0.64$ mag for the RR Lyrae stars as suggested by \citet{iorio_2019} and \citet{vasiliev_2021}, the dispersion in DM would not vary, but the bias would increase to 0.097 mag. It should be noted here that the trend is similar for the BHBs and RGBs of the SEGUE sample. However, for the BHB, the Sgr DM variation that we fitted on the RRLyrae leads to a scatter in absolute magnitude of the BHBs of 0.3 mag, instead of the typical scatter of  0.1 mag found by \citet{deason_2011}. This is again a consequence of the scatter of the fitted DM variation in the RRLyrae, of the intrisinc scatter of the Sgr stream, and of the uncertainties on our  estimation of spectro-photometric distances.

\begin{table*}
\centering
\label{param_pot}
\begin{tabular}{|l|l|l|l|l|}
\hline
\multicolumn{1}{|l|}{{\bf Disc}} &
  \multicolumn{1}{|c|}{$\Sigma_0$ (M$_\odot$.kpc$^{-2}$)} &
  \multicolumn{1}{c|}{R$_d$ (kpc)} &
  \multicolumn{1}{c|}{z$_d$ (kpc)} &
  \multicolumn{1}{c|}{R$_{m}$ (kpc)} \\
\hline
Thin & $1.4\times10^9$ & 2.3 & 0.18 & 0.0 \\
Thick & $1.2\times10^8$ & 2.3 & 1.0 & 0.0 \\
ISM & $2.2\times10^8$ & 4.6 & 0.04 & 4.6\\
\hline
\end{tabular}
\\
\begin{tabular}{|l|l|l|l|l|l|l|}
\hline
\multicolumn{1}{|l|}{{\bf Halo}} &
  \multicolumn{1}{|c|}{$\rho_0$ (M$_\odot$.kpc$^{-3}$)} &
  \multicolumn{1}{c|}{r$_0$ (kpc)} &
  \multicolumn{1}{c|}{r$_t$ (kpc)} &
  \multicolumn{1}{c|}{q} &
  \multicolumn{1}{c|}{$\gamma$}&
  \multicolumn{1}{c|}{$\beta$}  \\
\hline
Bulge & $4.8\times10^8$ & 0.6 & 1.1 & 0.6 & 1.8 & 1.8 \\
DM halo & $9.9\times10^6$ & 17.0 & 207 & 0.82 & 1 & 3\\
\hline
\end{tabular}
\caption{Parameters of the {\sc GalPot} potential listed here, composed of three disc components corresponding to the thin and thick discs and the interstellar medium, and two halo components corresponding to the bulge and the dark matter halo. The notation is the same as in \citet{dehnen_1998}.}
\end{table*}

\begin{figure*}
\centering
  \includegraphics[angle=0, viewport= 0 0 720 1275,clip,width=12cm]{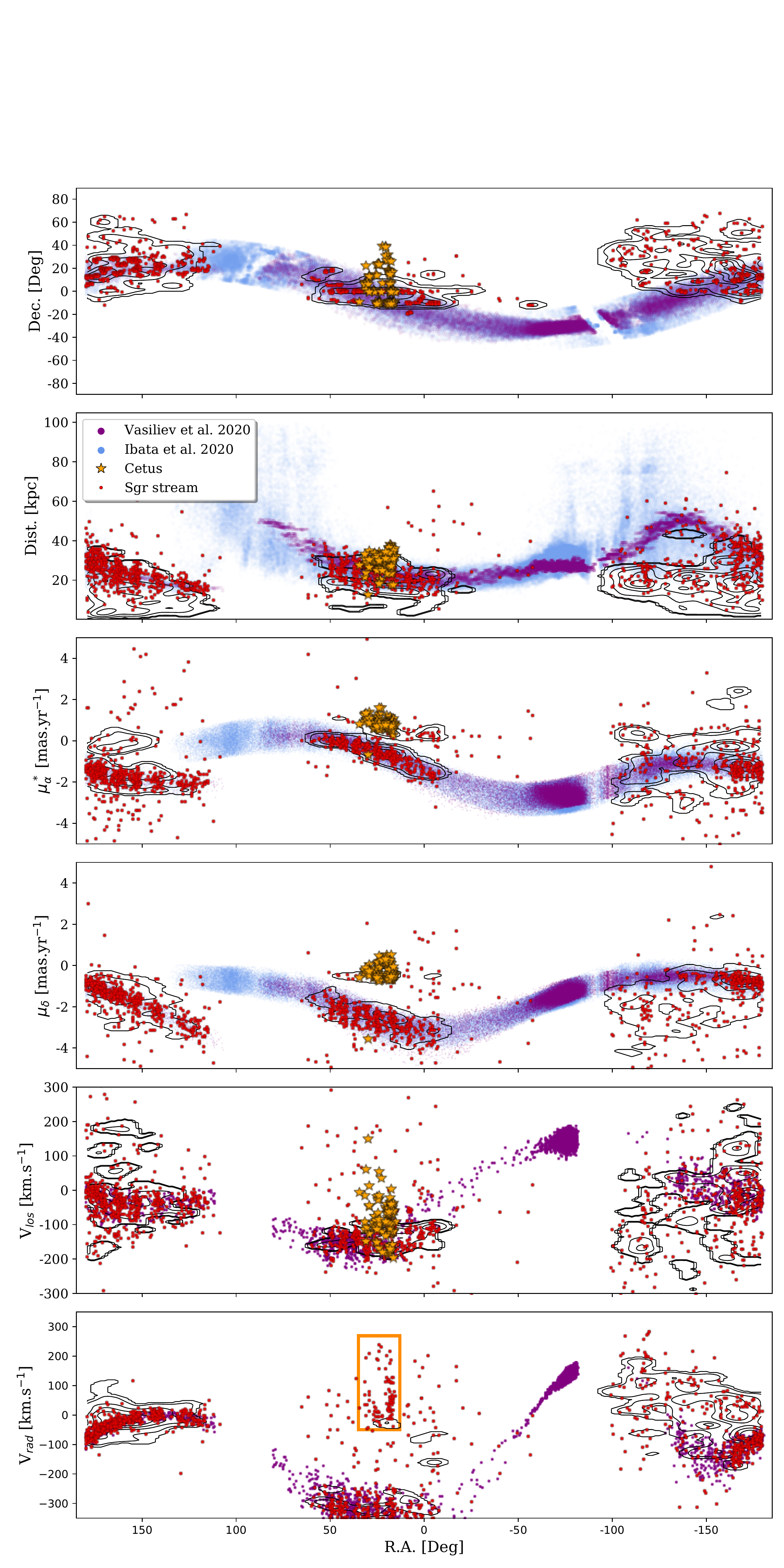}
   \caption{Position on the sky, heliocentric distance, proper motions, line-of-sight, and Galactocentric radial velocities of the stars from the structure that we identified in Fig.~\ref{phase-space} (red symbols). The light-blue and purple points are the stars identified as part of the Sagittarius stream by \citet{ibata_2020} and \citet{vasiliev_2021}, respectively. In the lower panel, another structure, highlighted by the orange rectangle, stands out around R.A.$\sim 25 \degr$, and is made up of stars that have clearly distinct galactocentric radial velocities  from both the Sagittarius stream and the contaminating background stars  (mostly originating from the Gaia-Enceladus-Sausage) shown by the black contours in each panel. The stars from this structure, which appears to be the Cetus stream, are shown by the orange stars in the other panels; they also form a distinctly visible structure in $\mu_{\alpha,*}$ and $\mu_{\delta,*}$} 
\label{Sgr_pos}
\end{figure*}

\section{Scientific application: Exploration of the integral of motion-space}\label{Sec_phase_space}

The catalogue of distances obtained in Sect.~\ref{sec:method} enables the determination of the full six dynamical dimensions for a large number of distant stars with SEGUE spectroscopic and {\it Gaia} EDR3 proper motion measurements, without being restricted to a few specific stellar populations such as BHB, RC, or RR Lyrae stars.  This allows exploration of the structures present in the outer disc--stellar halo integral of motion space (also called phase-space), similar to that recently conducted by \citet{naidu_2020} with the H3 survey \citep{conroy_2019}. The integrals of motion are particularly useful for identifying past accretion events, because the stars originating from the same object are still clustered in phase-space, even several gigayears after the spatial coherence of the progenitor stellar component has been lost 
\citep{helmi_1999,jean-baptiste_2016}.

\begin{figure}
\centering
  \includegraphics[angle=0, clip,width=8.5cm]{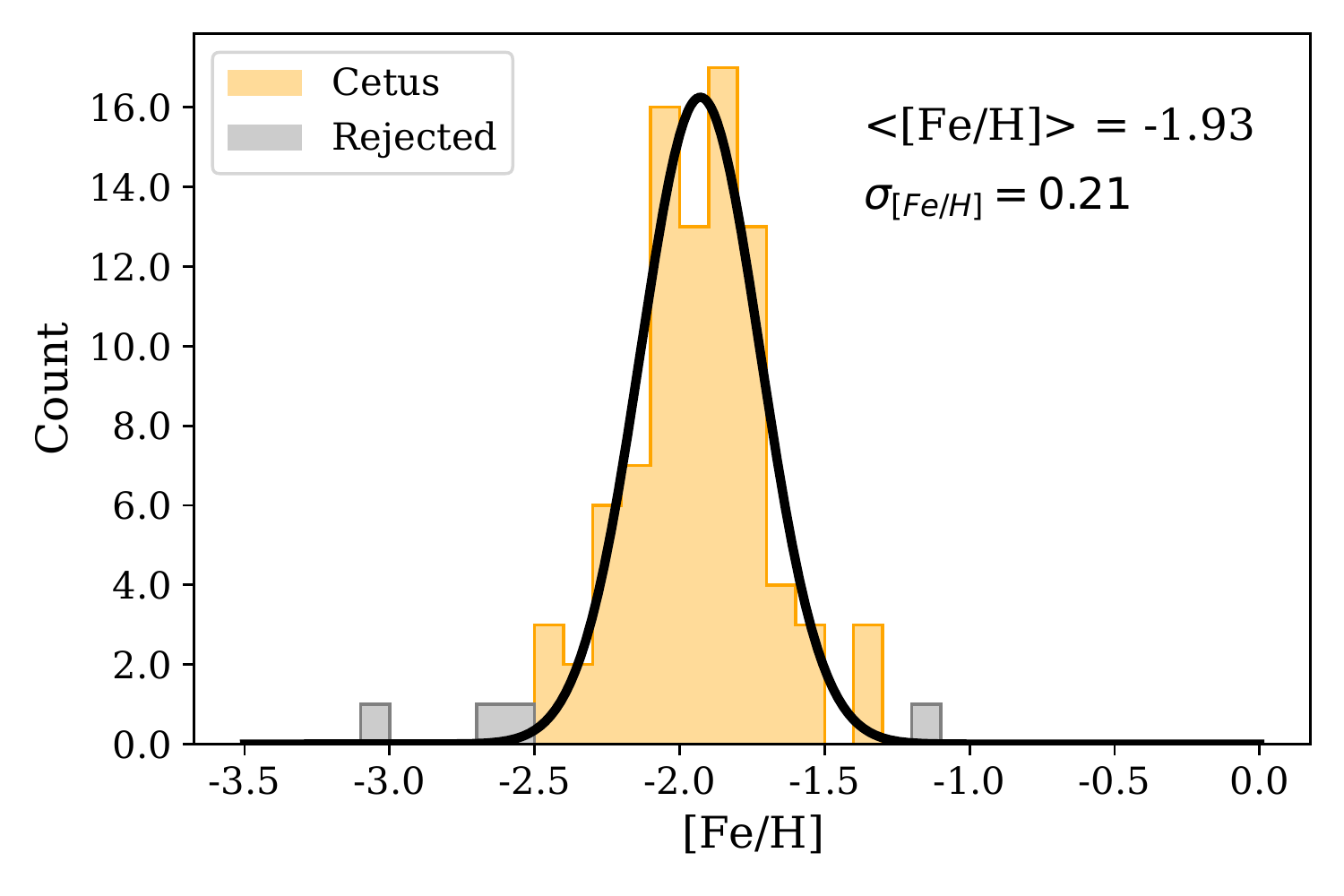}
    \includegraphics[angle=0, clip,width=8.5cm]{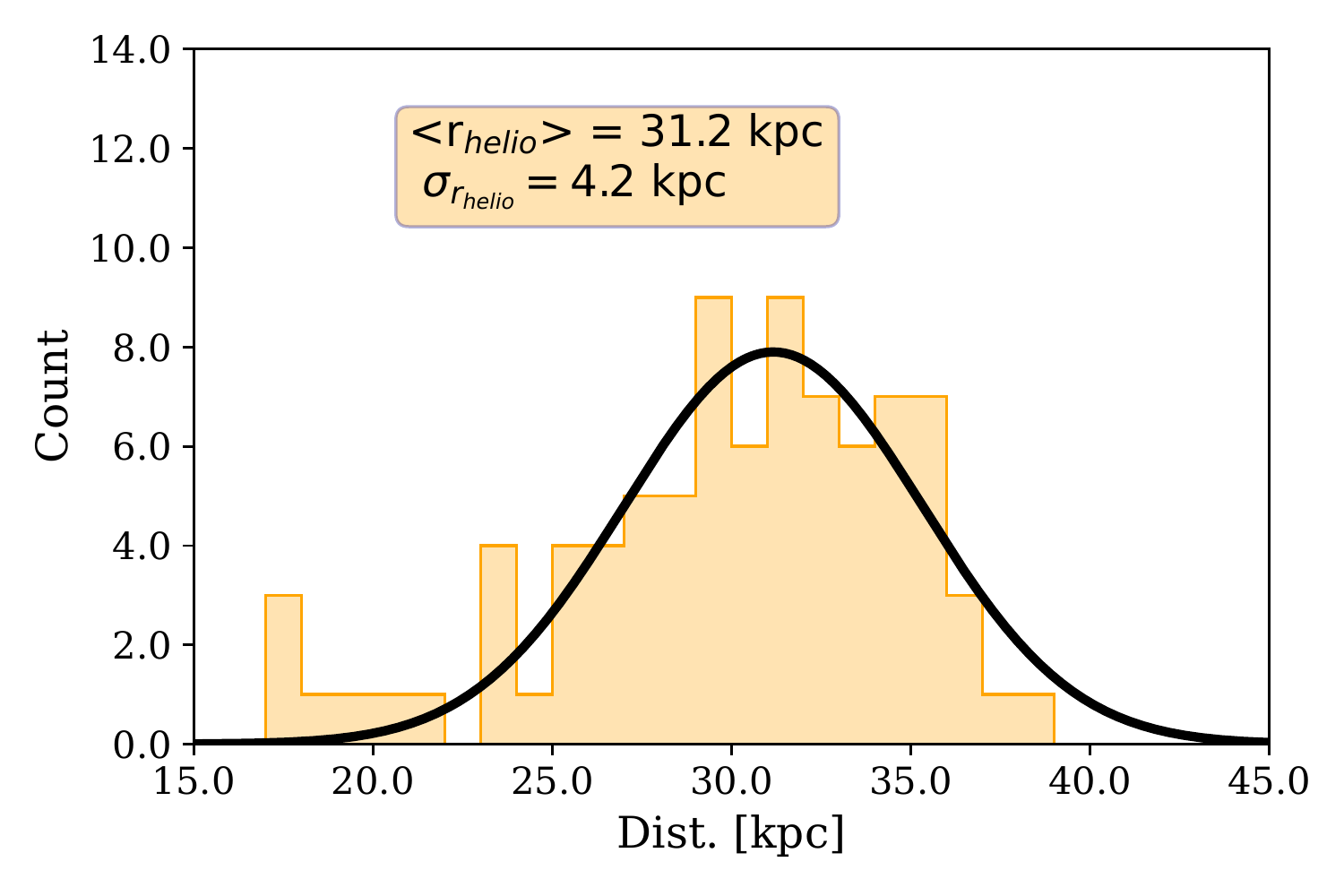}
   \caption{{\it Upper panel:} Metallicity distribution of the Cetus stream spectroscopic sample. The grey histogram shows the metallicity of the four stars that were rejected by the sigma-clipping method. {\it Lower panel:} Distance  distribution of the Cetus stream spectroscopic sample. The distribution is clearly asymmetric, with a tail towards shorter distances, which could be an indication of a distance gradient along the stream.} 
\label{fig:feh_dist_cetus}
\end{figure}

For the stars of the cross-matched SEGUE-PS1-EDR3 catalogue, we obtain the Cartesian Galactocentric positions and velocities in a right-handed system, assuming that the circular velocity rotation at the solar radius \citep[R$_\odot$=8.129 kpc, Z$_\odot$=20.4 pc][]{gravitycollaboration_2018} is  $v_0=229.0$ km.s$^{-1}$ \citep{eilers_2019}, with the Sun peculiar motions from \citet{schonrich_2010} ($U_\odot$, $V_\odot$, $W_\odot$) = (11.1, 12.24, 7.25) km.s$^{-1}$. The corresponding integral-of-motion quantities are computed using the St\"aeckel approximation method from the {\sc Agama} Python package \citep{vasiliev_2018}. The potential used for this computation is based on the {\sc GalPot} potential \citep{dehnen_1998,mcmillan_2017}, which is composed of three exponential discs, representing the stellar thin and thick discs, the interstellar medium, the bulge, and the dark matter halo, whose parameters, listed in Table \ref{param_pot}, have been slightly adjusted to fit the rotation curve observed by \citet{eilers_2019}, as shown in Figure \ref{rotcurv}. 

Figure~\ref{phase-space} (left) shows the distribution of the azimuthal ($J_{\phi}$)  and vertical ($J_z$) actions of the SEGUE stars. One can see a substructure around $J_\phi \simeq -3000$ kpc km s$^{-1}$ at low $J_z$, which corresponds to the Galactic disc, as confirmed by the corresponding position on the energy-vertical angular momentum\footnote{The vertical angular momentum is equal to the azimutal action in an asymmetric potential, as used here.} diagram (Fig.~\ref{phase-space}, right) close to a circular orbit (represented by the green line). In this last panel, one can also see a structure around L$_z\sim 2000$ kpc.km.s$^{-1}$ and E$\sim-0.7 \times 10^5$ km$^2$.s$^{-2}$, which corresponds to Sequoia \citep{myeong_2018a,myeong_2018b,myeong_2019}. In addition to the disc, an elongated overdensity in the azimuthal and vertical action plane is clearly visible (highlighted
by the red ellipse; see the right panel of Fig.~\ref{phase-space} for the location of the stars in E-L$_z$). The large majority of the stars forming this structure are part of the Sagittarius stream \citep{ibata_1994}, as is visible in Fig.~\ref{Sgr_pos}, where we compare their positions, heliocentric distance, proper motions, line-of-sight, and Galactocentric radial velocities to those of the Sgr stream identified previously by \citet{ibata_2020} and \citet{vasiliev_2021}. 

However, as we can see from the different panels of Figure~\ref{Sgr_pos}, aside from the Sgr stream, our action-space selection encompasses stars from the foreground and background \citep[which includes other structures too, like the Gaia-Enceladus-Sausage; see][]{naidu_2020} and a clear substructure at R.A.$=10-35 \degr$, which stands out in radial velocity and in the two proper-motion components. To gauge the contribution from foreground and background stars, we overlay on Fig.~\ref{Sgr_pos} (shown by the black contour) the distribution of the stars lying at the edge of the action-space selected area (i.e. between the red and blue ellipses on Fig.~\ref{phase-space}): here one can see stars mostly belonging to the Gaia-Enceladus-Sausage \citep[see e.g.][]{naidu_2020,myeong_2018}, but also a few stars with low $J_z$ belonging to the Sgr stream not included in our ellipsoid selection. Nevertheless, the group of 91 stars between R.A.$=10-35 \degr$ and Galactocentric radial velocity between V$_{rad}=-50$ and 270 km.s$^{-1}$ (within the orange box in the lower panel of Fig.~\ref{Sgr_pos} and highlighted with star symbols in the others) have a clearly different distribution from both the stars of the Sgr stream and the contamination. Comparing the position on the sky and the distances of those stars to previous works \citep{newberg_2009,koposov_2012,yam_2013}, we conclude that those stars are part of the Cetus stellar stream. We refer to those stars as the Cetus stream spectroscopic sample in the remainder of the paper. 

Despite sharing similar positions in energy, and vertical and azimuthal actions, the spatial track of the Cetus stream differs from that of the Sgr stream by $\sim 60 \degr$, as already mentioned in these previous studies. This example clearly shows the limitation of the identification of stellar structures with only the energy--angular momentum diagram or in actions-space, without considering the action-angles, because these methods do not use the full 6D information available. 

The metallicity distribution (obtained using the {\sc Fehadop} parameter) of the 91 stars in the Cetus stream spectroscopic sample is shown in the upper panel of Fig.~\ref{fig:feh_dist_cetus}. The mean metallicity is  [Fe/H]=-1.93 with a dispersion  $\sigma_{[Fe/H]}=$0.21 dex, consistent with the values found by \citet{yam_2013} and \citet{yuan_2019}. These values are insensitive to being derived from the full sample of 91 stars or by applying a 3-$\sigma$ clipping, which removes four stars.

The mean heliocentric distance we obtain for the Cetus stream spectroscopic sample is  31.2 kpc with a total dispersion of 4.2 kpc, including the distance uncertainties, which are consistent with previous measurements \citep{newberg_2009,koposov_2012,yam_2013,yuan_2019} made in the same part of the sky, and with various stellar tracers (RGBs, BHBs, and K-giants). However, the distance distribution (Fig.~\ref{fig:feh_dist_cetus}, lower panel) appears asymmetric, with a tail pointing towards shorter distances ($\sim 20$ kpc), which could be an indication of a distance variation along the stream. Unfortunately, the spectroscopic Cetus sample is not sufficient to measure a distance gradient. In the following section, we expand the search for Cetus stream member stars using 5D information from {\it Gaia} EDR3 data.
\begin{figure}
\centering
  \includegraphics[angle=0, clip,width=8.5cm]{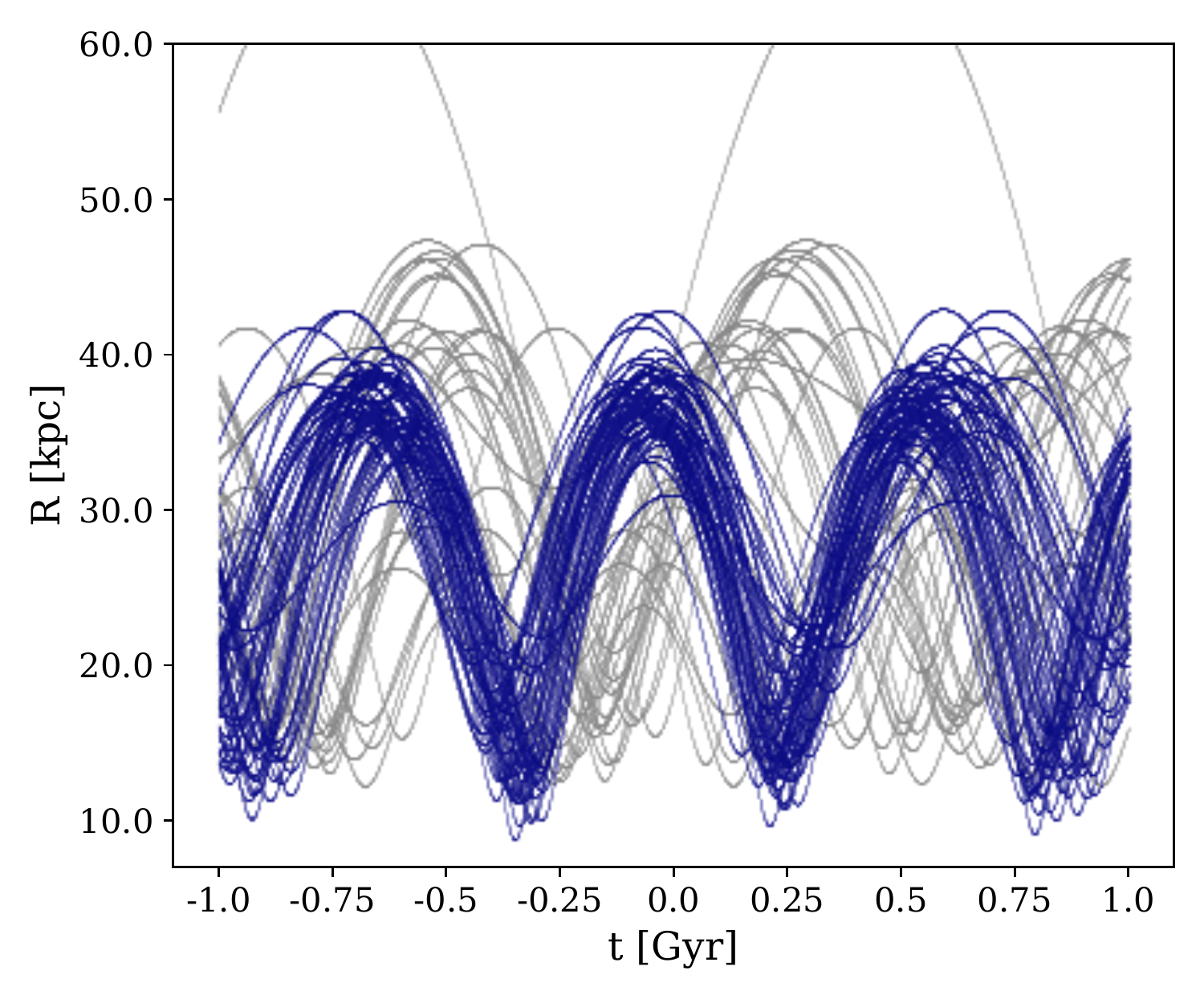}
   \caption{Orbits of the Cetus stream spectroscopic sample stars integrated over 1 Gyr backward and forward. The blue orbits are the ones used to find the natural coordinates of the Cetus stellar stream.}
\label{fig:orbits}
\end{figure}

\begin{figure*}
\centering
  \includegraphics[angle=0, viewport= 0 170 720 550,clip,width=14cm]{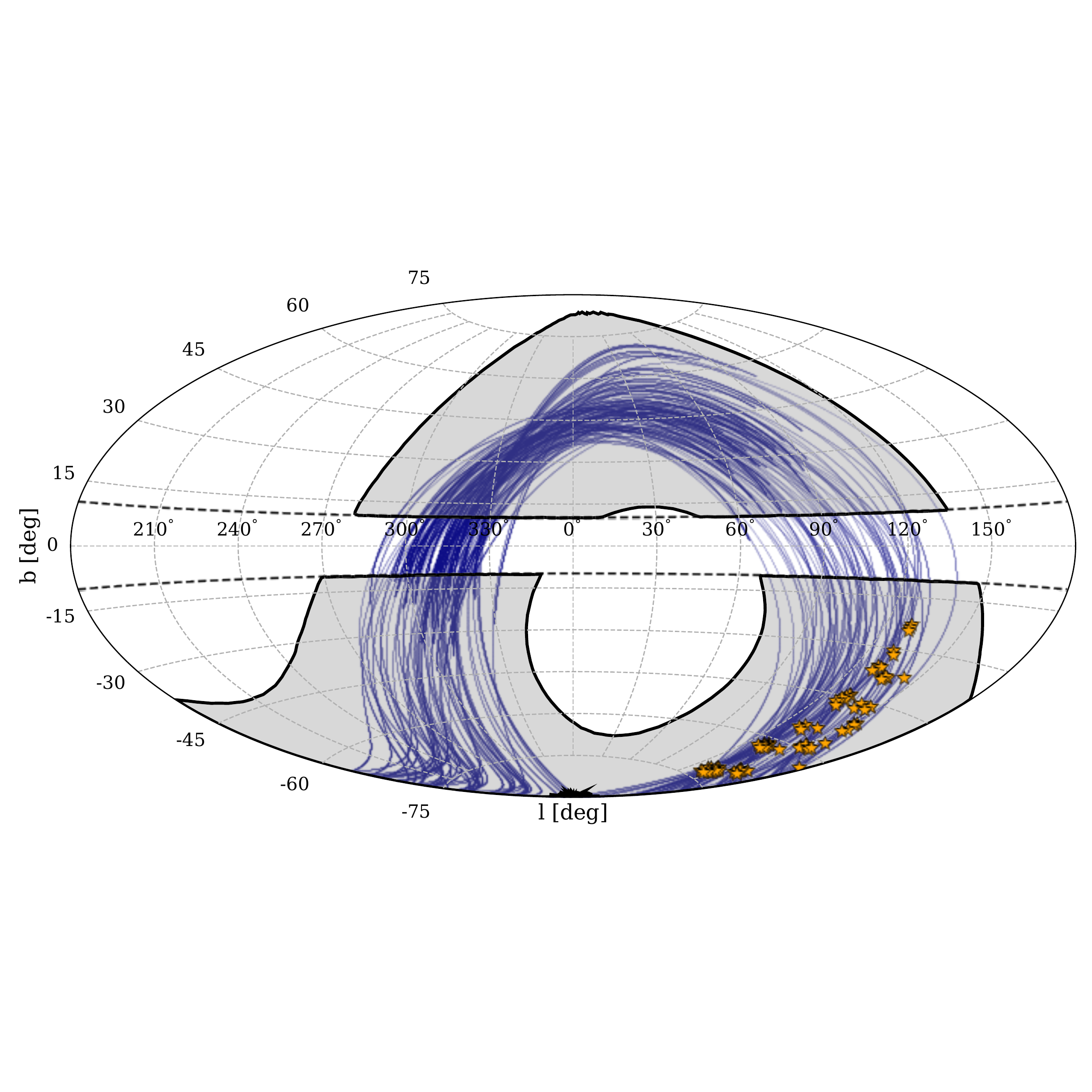}
   \caption{Aitoff projection of the sky in Galactic coordinates; the area used to search for the Cetus stream is represented in grey. The region near the Galactic disc ($|b|<10 \degr$), within the dashed lines, is excluded from the search because of the high extinction and the large number of contaminant disc stars. The blue lines correspond to the orbits of the stars from the Cetus spectroscopic sample used to find the plane of the Cetus stream. The current location of these stars is indicated by the orange stars.} 
\label{fig:sel}
\end{figure*}

\section{The Cetus stream in {\it Gaia} EDR3} \label{sec:cetus_erd3}

In this section, we expand the search for Cetus member stars to the 5D {\it Gaia} EDR3 sample in order to identify the edges of the stream, measure its  distance gradient, and estimate the stellar mass of its progenitor. To this aim, we first use the 6D information from the Cetus stream spectroscopic sample to guide our search; specifically, we integrate the orbits of these stars to define the area of the sky likely to contain its tidal debris (Sect.~\ref{sec:plane}). The search for Cetus stream member stars is first carried out on BHB stars, whose absolute magnitude (and therefore distance) can be derived from photometric colours with good precision.  For example, such a relation exists for SDSS $(g-r)$ \citep{deason_2011}; as no such relation is currently available for the {\it Gaia} bands, and we do not want to be limited by the SDSS coverage, we first provide a re-calibration of the \citet{deason_2011} relation to the {\it Gaia} EDR3 photometric system (Sect.~\ref{sec:dist_BHB}) and then use the likely BHBs identified in {\it Gaia} EDR3 data to analyse the Cetus stream, and to define its extent, distance gradient, and track on the sky (Sect.~\ref{sec:BHB_cetus}). In Sect.~\ref{sec:mass},  the properties derived in \ref{sec:BHB_cetus} are used to include stars in other evolutionary phases, assign probabilities of membership, and estimate the mass of the progenitor.

\begin{figure}
\centering
  \includegraphics[angle=0,clip,width=8.5cm]{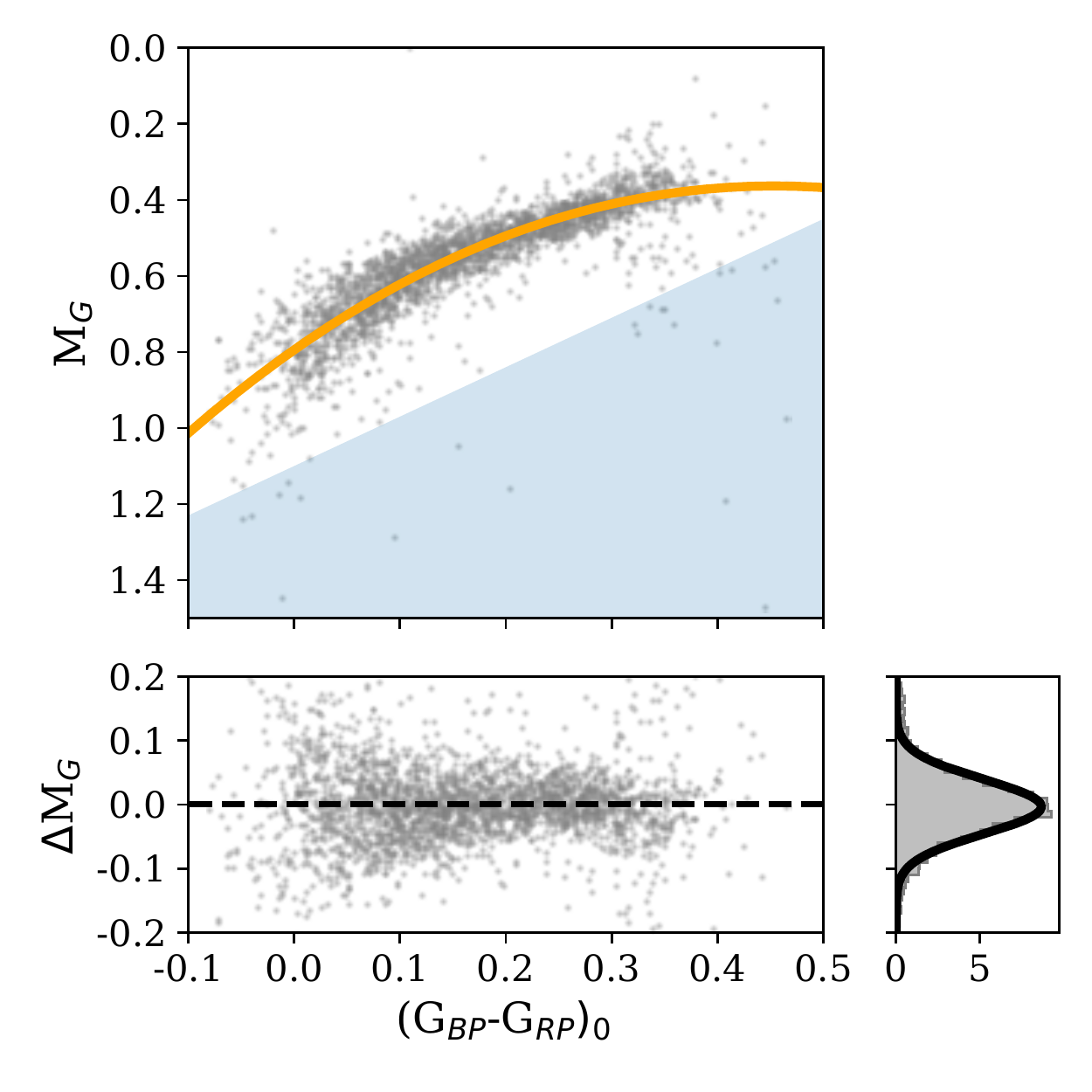}
   \caption{Colour--absolute magnitude relation for BHB stars in the {\it Gaia} EDR3 photometric system. The orange line corresponds to the relation found in this work (Equation~\ref{eq:dist_BHB}) by fitting a polynomial on the BHB population by \citet{xue_2011} (grey points). The stars in the blue area are excluded from the fit as they are likely misidentified as BHB. The lower panels show the residuals of the absolute magnitude to the fitted relation.}
\label{fig:BHB_dist}
\end{figure}

\subsection{The orbital plane of the Cetus stream} \label{sec:plane}

The track of stellar streams emanating from the disruption of globular clusters or small dwarf galaxies is well approximated by the orbital path of their progenitor (e.g.  \citealt{malhan_2018a} and \citealt{ibata_2021} used this method to detect new stellar streams in {\it Gaia} data). Recent simulations of \citet{chang_2020} show that the track of the Cetus stream follows the orbit of its progenitor over two wraps, even if it originated from a dwarf galaxy with a $2 \times 10^9$ M$_\odot$ total mass. Therefore, it is possible to use the orbit of the stars of the Cetus stream spectroscopic sample to define its orbital plane, and thus the area of the sky where tidal debris is to be searched for. 

The orbits were integrated backward and forward over 1 Gyr in the same MW potential used in the previous section. The variation in Galactocentric distance as a function of time is shown in Fig.~\ref{fig:orbits}. Over the 87 orbits, one star is clearly not related to the Cetus stream, with a completely different  apocenter from all the others. The orbits of the majority of the other stars have a similar common behaviour (56/86, highlighted in blue in the figure), with a pericenter of $\sim 15$ kpc, an apocenter at $\sim 37$ kpc, an orbital inclination of $\sim 65 \degr$ , and an orbital period of $\simeq 600$ Myr, which is slightly lower than the value of 700 Myr found by \citet{yam_2013}, betraying a more than probable common origin. Therefore, only two periods, one forward and one backward (i.e. between $\pm 600$ Myr) of the 56 orbits with a common behaviour are used to define the orbital plane of the Cetus stream. The differences of most of the other 30 orbits from this common behaviour can be explained by the fact that these stars are not part of the Cetus stream. Indeed, the current galactocentric radial velocity of 14 of them exceeds 150 km.s$^{-1}$, contrary to the stars showing a common orbital behaviour, whose galactocentric radial velocities range from 0 to 150 km.s$^{-1}$. For ten others, the  current distance ranges between 18 kpc and 25 kpc, while the current distance of the stars with a common orbital behaviour is between 25 kpc and 40 kpc. For the six others, the deviating orbital behaviour could be a consequence of the combined uncertainties on the observables: proper motions (PM), distance, and radial velocity). We note that it is also possible that the departure from the common behaviour of all these stars is caused by the simple assumption we made that the Cetus stream is a fully coherent structure disrupted in an adiabatic process, which neglects for example the perturbations generated by a massive Large Magellanic Cloud (LMC). Interestingly, removing those 30 stars does not change the mean metallicity that we find for the Cetus stream of [Fe/H]$=-1.93 \pm{0.21}$, while the mean heliocentric distance increases slightly and the dispersion is reduced (from 31.2 to 32.0 kpc and from 4.2 to 2.8 kpc, respectively).

Although the `natural' coordinates of a stellar stream are usually expressed using a great circle rotation \citep[i.e.][]{ibata_2001,koposov_2010,koposov_2019,shipp_2019}, in the case of the Cetus stream, we find that a small circle is better suited. Indeed, in a Galactocentric frame, the projected coordinates will follow a great circle, but this is no longer the case if the observer is located on Earth, and a constant offset is visible in the coordinate perpendicular to the stream. We find that the `natural' coordinates of the Cetus stream ($\phi_1$, $\phi_2$) are well expressed by a small circle with a pole centred on (R.A., Decl.)=($125.1809832\degr$, $15.91290743\degr$) and an offset to the great circle of $14.36 \degr$. The transformation of coordinates is performed with the {\sc GreatCircleICRSFrame} class present in the {\sc Gala} package \citep{price-whelan_2017}, with $\phi_1$, $\phi_2$)=(0,0)  corresponding to (R.A., Decl.)=(22.11454259\degr, -6.7038421\degr), where we think the centre of the progenitor of the Cetus stream is located (see Section~\ref{sec:BHB_cetus}). 

Hereafter, we limit the search of stars belonging to the Cetus stream to the region between $|\phi_2|<35 \degr$, because this latter contains the 56 orbits used to define the stream orbital plane. Moreover, we exclude the region at  $|b|<10\degr$ because of the high extinction and the large number of disc stars. The corresponding search area is highlighted in grey in Fig.~\ref{fig:sel}.

\begin{figure*}
\centering
  \includegraphics[angle=0,clip,width=17.5cm]{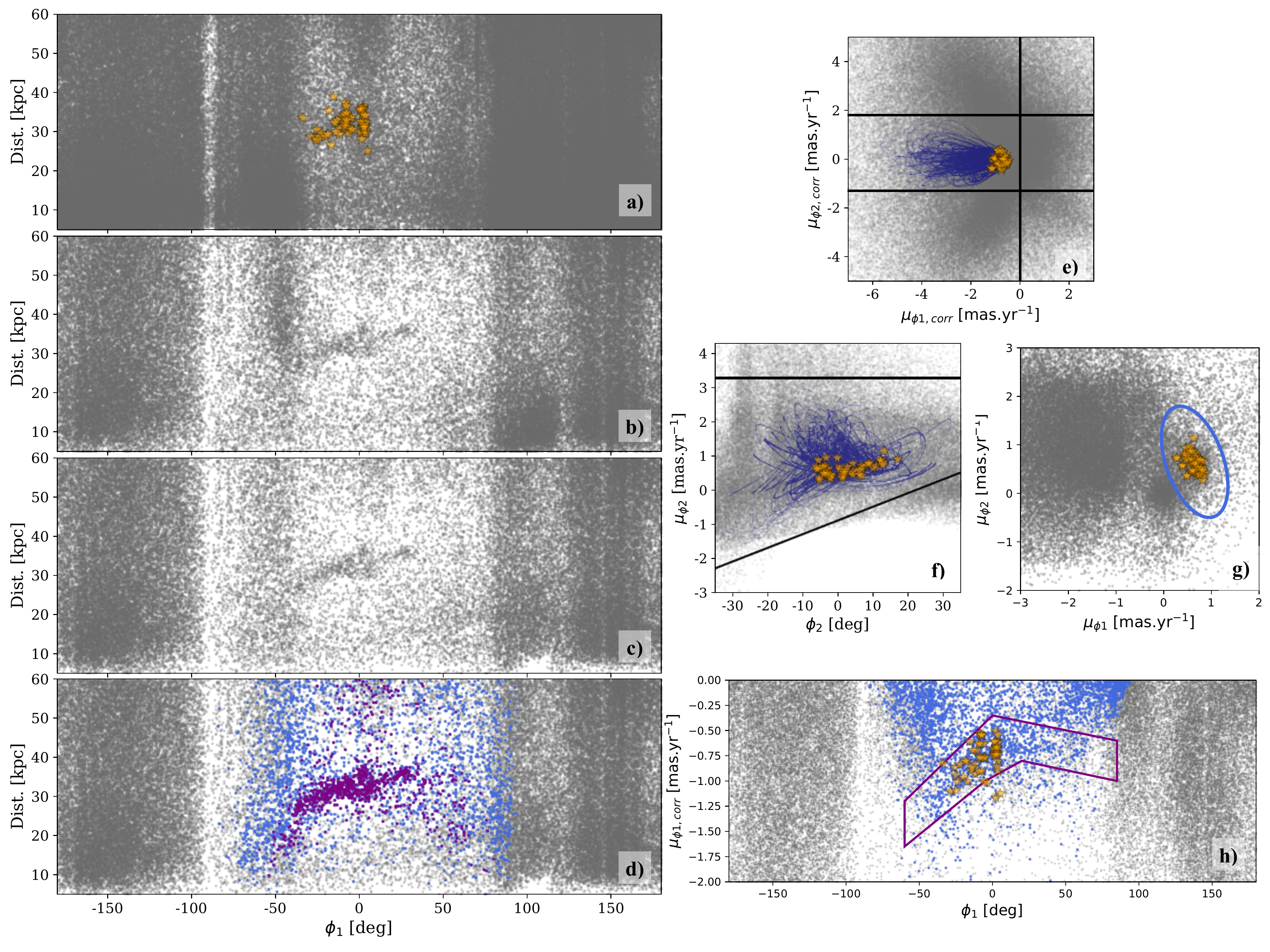}
   \caption{Distance distribution of selected BHB stars through a succession of increasingly restrictive kinematic selections. {\it Panel (a)} shows the distance distribution of BHB stars prior to any kinematic selections, where the orange stars show the positions of the 56 spectroscopic stars whose orbits have been used to define the orbital plane of the Cetus stream. {\it Panel (b)} shows the same distribution for stars after a first broad kinematic cut performed in the corrected proper motion, illustrated by the black lines in {\it panel (e)}, where the blue lines show the orbits of the spectroscopic stars used to define this cut. {\it Panel (c)} shows the distance distribution of BHB stars after an additional wide selection based on their non-corrected proper motions, as shown in {\it panel (f)} (see Section~\ref{sec:dist_BHB}). {\it Panel (d)} shows the distribution of the BHBs of the Cetus stream (in purple) after two additional more restrictive selections to remove the contamination of foreground and background stars performed in the non-corrected proper motion space shown in {\it panel (g)} and in the position-corrected parallel proper motion space shown in {\it panel (h)}. In all panels, $\phi_1$ and $\phi_2$ refer to the `natural'  coordinates  of  the  Cetus  stream (see Section~\ref{sec:plane}).} 
\label{fig:BHB}
\end{figure*}

\subsection{Tracing the Cetus stream with BHB stars}

 Blue horizontal branch  stars are very often used to estimate the distance of substructures because they are relatively bright stars and their absolute magnitude is constant to first approximation \citep[M$_g \sim 0.5$ in the SDSS $g$-band; ][]{deason_2011}. Therefore, we first use this stellar population to investigate the extent of the Cetus stellar stream and to measure the distance gradient along it.

\subsubsection{BHBs distance calibrated for {\it Gaia} EDR3} \label{sec:dist_BHB}

\citet{deason_2011} found that the absolute magnitude of BHB stars, and therefore their heliocentric distances, are a function of their colour, with a narrow scatter around the relation derived by these authors ($\sim 0.1$ mag). This relation was calibrated for the SDSS photometric system; however, SDSS only covers a fraction of the orbital plane of the Cetus stream. In order to exploit the full sky coverage afforded by {\it Gaia}, we re-calibrate the \citet{deason_2011} relation in the {\it Gaia} EDR3 photometric system. 

To this aim, we used the sample of BHBs  identified spectroscopically by \citet{xue_2008} in SEGUE, adopt their SDSS photometry to derive their DM using the relation by \citet{deason_2011}, and use it to compute their absolute magnitude in the {\it Gaia} $G$-band, $M_{\rm G,BHB}$. As shown in  Fig.~\ref{fig:BHB_dist}, $M_{\rm G,BHB}$ can be expressed as:
\begin{equation}
\begin{array}{ll}
    M_{G,BHB}=& -0.266(G_{BP}-G_{RP})_0^3+ 2.335  (G_{BP}-G_{RP})_0^2\\    &-1.955 (G_{BP}-G_{RP})_0+0.794 \, .
\end{array}
\label{eq:dist_BHB}
\end{equation}
Here we excluded the stars with $M_G>-1.3 (G_{BP}-G_{RP})_0 +1.1$, because they clearly deviate from the general trend, possibly because the \citet{xue_2008} BHB sample is not 100\% pure \citep[see][]{starkenburg_2019}. It has to be noted that 2160  of the 2362 stars used are located in the range of validity of the \citet{deason_2011} relation, namely $-0.25 \leq(g-r)_0 \leq 0.0$.

The scatter between the absolute magnitude estimated by our re-calibration and that by  \citet{deason_2011} is  $\sigma_{\Delta M_G} = 0.04$ mag, which is smaller than the intrinsic precision of 0.1 mag of the \citet{deason_2011} relation, and with a negligible offset  ($<0.005$ mag). This means that our method has a relative distance precision similar to the \citet{deason_2011} calibration, i.e. 5\%. 

\subsubsection{Identification of BHB candidates along the Cetus stream} \label{sec:BHB_cetus}

In order to search for BHB candidates along the Cetus stream, we selected {\it Gaia} EDR3 sources with G$_{BP}$-G$_{RP})_0 \leq 0.5$ located in the plane of the Cetus stream ($|\phi_2|<35\degr$). We removed sources that had the duplicate flag on or a spurious astrometric solution (with the re-normalised unit weight error, {\sc RUWE} $> 1.4$, Lindegren, document {\sc GAIA-C3-TN-LU-L-124-01}), and those located in high-extinction areas (E(B-V)$>$0.3) and those with high uncertainties on their parallax measurement ($\varpi > 0.4$ mas).   

As the Cetus stream has a pericenter of $\sim 10$ kpc (Fig.~\ref{fig:orbits}), we applied a generous cut on the parallax such that stars with $\varpi-3 \delta \varpi < 1/5$ mas are kept; this removes a large portion of disc contamination. To limit contamination, we also removed stars located within: 15 times the  half-light radius of any globular clusters around the MW \citep[using the values given by][]{harris_2010}; 5 times the half-light radius of the dwarf galaxies listed in the catalogue of \citet{mcconnachie_2012}, an ellipse centred on M31 with a semi-major axis of $3\degr$ and a semi-minor axis of $1.5\degr$ oriented at $38\degr$ \citep{devaucouleurs_1958} and within the inner $1\degr$ of M33; an ellipse centred on the LMC with semi-major and minor axes of $26\degr$ and $11.6\degr$ respectively, and in an ellipse centred on the SMC with a semi-major axis of $20\degr$ and a semi-minor axis of $8\degr$. For M31, M33, the LMC, and the SMC, these values have been chosen by visual inspection.

The distance distribution of the selected blue point sources is shown in Fig.~\ref{fig:BHB}(a) as a function of longitude along the Cetus stream orbital plane, $\phi_1$. These contain  not only BHB stars, but also contamination from blue stragglers (BSs), white dwarfs (WDs), young MS turn-off stars (y-MSTOs), and distant quasi-stellar objects \citep[QSOs; e.g.][]{sirko_2004,vickers_2012,fukushima_2018}. We removed the known QSOs compiled by \citet{liao_2019}. As for removing the majority of the stellar contaminants, we  perform a series of gradually more restrictive kinematic selections. Such a process allows in a first step to detect the full extent of the stream without cutting its edges, by performing a very restrictive selection, and then to clean the sample in a second step.
Stars in a stream generally do not have large motions in the direction perpendicular to the orbit, unless there are strong perturbers \citep[e.g.][]{erkal_2019}, and have a common direction of movement parallel to the orbit. For these reasons, based on the orbits of the Cetus stream spectroscopic sample, we performed a first  broad kinematic selection, which is visible in Fig.~\ref{fig:BHB}(e), such as ($\mu_{\phi 1,corr}<0.0\ \mathrm{mas}.\mathrm{yr}^{-1}$ and $-1.3\ \mathrm{mas}.\mathrm{yr}^{-1} <\mu_{\phi 2,corr}<1.8\ \mathrm{mas}.\mathrm{yr}^{-1}$), where $\mu_{\phi 1,corr}$ and $\mu_{\phi 2,corr}$ are the proper motions along the Cetus coordinates, corrected for the Sun reflex motion using the individual distances estimated from Equation~\ref{eq:dist_BHB} and the solar motion used in Section~\ref{Sec_phase_space}. As visible in Fig.~\ref{fig:BHB}(b), we can already clearly identify the stream structure between $-40 \degr < \phi_1 < 40 \degr$. This wide generous selection still includes contamination, such as  background/foreground BHBs from the Galactic discs and stellar halo, as well as blue stars misidentified as BHBs. 

As distances (and corrected proper motions) for misidentified BHBs will be incorrect, we also look into non-corrected proper motions. Based on the orbits of the stars derived earlier, we see that a second broad selection, keeping stars with $3.3$ mas.yr$^{-1} >\mu_{phi,2}>0.04 \cdot \phi_2-0.9$ mas.yr$^{-1}$, as shown in Fig.~\ref{fig:BHB} (f), removes a small fraction ($\simeq 16\%$) of the sample, but improves the detection of the edges of the stream around $\phi_1 \simeq \pm 50 \degr$ (Fig.~\ref{fig:BHB} c). 
In equatorial coordinates, this region corresponds to an area located between $-10 \degr \lessapprox$ R.A.$ \lessapprox 50 \degr$ and $ -90 \degr  \lessapprox$ Decl.$ \lessapprox 55 \degr$, extending the previous detection of the Cetus stream \citep{newberg_2009,yam_2013,yuan_2019} over several tens of degrees in the equatorial Southern Hemisphere, overlapping with the position of the Palca overdensity discovered by \citet{shipp_2018}, as predicted recently by \citet{chang_2020}. Therefore, to avoid future confusions, we propose to rename the stream the Cetus-Palca stream.
However,  in neither Fig.~\ref{fig:BHB} (b) nor (c) do we see the trace of the northern (Galactic) counterpart detected by \citet{yuan_2019} in phase space, which should be located around $\phi_1 \sim 150 \degr$ ( $l \sim$50$^{\circ}$, $b \sim$50$^{\circ}$). The detection of clear edges to the Cetus-Palca stream, and the absence of broadening toward them in Fig.~\ref{fig:BHB} (c), may suggest that the northern detection of \citet{yuan_2019} is an artefact, or not related to the Cetus-Palca stream. However, this northern counterpart is expected to be spatially diffuse, and it may not appear as a structure distinguishable from the background or foreground in our selection.

To clean out the broadly selected sample, we applied two more restrictive selections, once again in kinematic space, because selecting in position and distance spaces could lead to cutting the stream before its actual edges. The first cut was done in (non-corrected) proper-motion space, as the Cetus-Palca stream stands out from most of the Milky Way contaminants, as highlighted by the blue ellipse in Fig.~\ref{fig:BHB} (g).  The stars within this ellipse are indicated by the blue points in Fig.~\ref{fig:BHB} (d) and (f). As this sample still contains a large amount of contaminants, mostly from the disc, we look at the position-parallel proper-motion space (Fig.~\ref{fig:BHB} h) to select the BHB stars that are part of the same structure as the stars from the spectroscopic sample. The final cleaned sample of BHB stars that are part of the Cetus-Palca stream is shown in purple in Fig.~\ref{fig:BHB} (e). We can clearly see the Cetus-Palca stream between $-40\degr \leq \phi_1 \leq 35\degr$. Around $\phi_1 =0 \degr$, the stream appears wider compared to the other sections, which could indicate the position where the leading and trailing arms meet, and so where the progenitor would likely be if it were completely disrupted. This motivated our choice for the location of the origin of the $\phi_1$ axis in Sect.~\ref{sec:plane}.

With the final BHB selection, one can see a clean, almost linear distance gradient along the Cetus-Palca stream, with a distance varying between $\sim 25$ kpc at the edge of the leading arm ($\phi_1 \simeq -40 \degr$) to $\sim 37$ kpc at the edge of the trailing arm ($phi_1 \simeq 35 \degr$). This explains the asymmetric distribution in the distance found with the Cetus-Palca stream spectroscopic sample, because SEGUE covers only the leading arm ($\phi_1<0 \degr$) of the Cetus-Palca stream.

\begin{figure*}
\centering
  \includegraphics[angle=0, viewport= 80 50 910 810,clip,width=17.5cm]{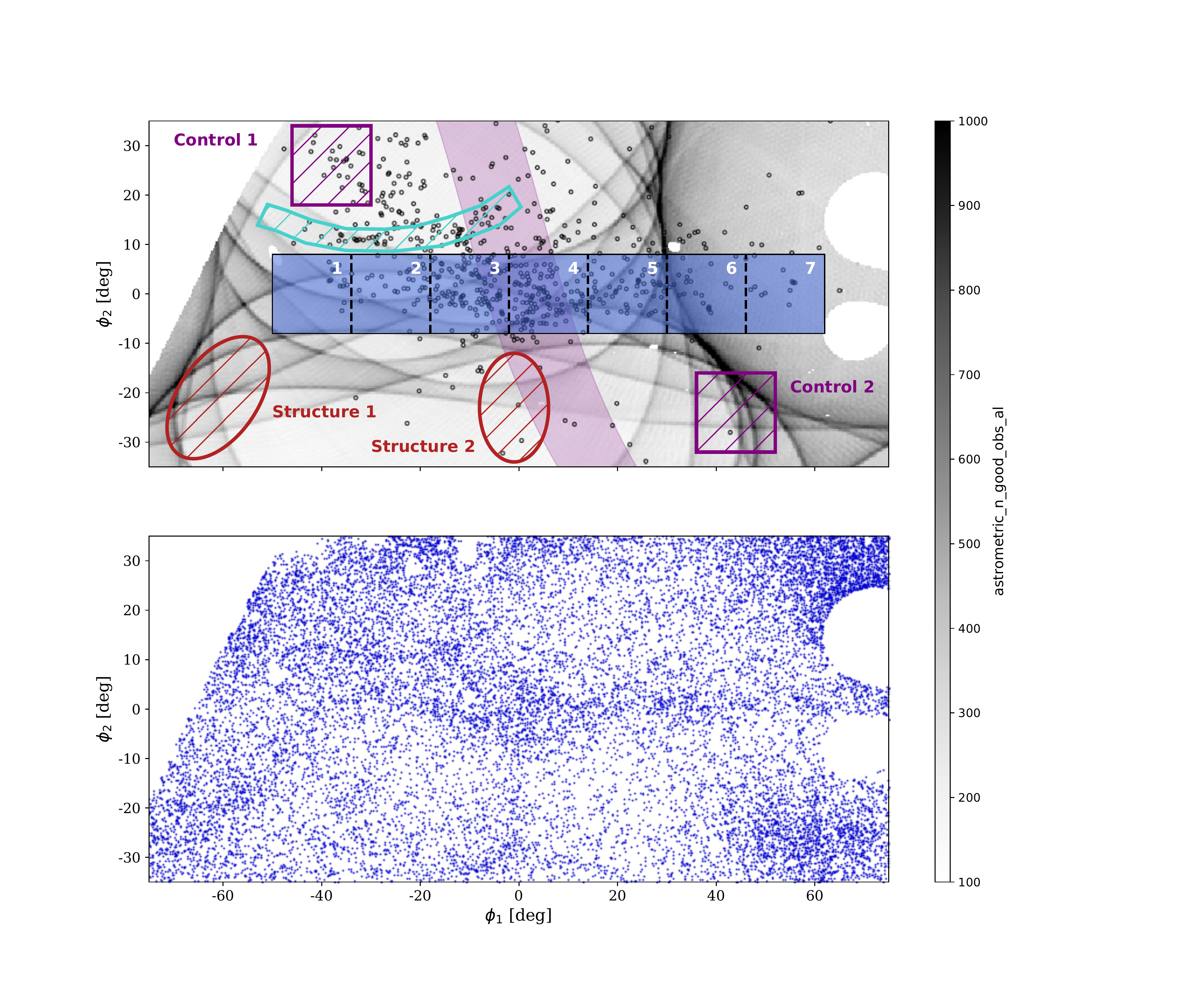}
   \caption{{\it Lower panel:} Position of the stars with $P_{Cetus}\geq 0.2$. {Upper panel:} The black circles show the position of the BHB of Cetus-Palca identified in Section~\ref{sec:BHB_cetus}, and the yellow stars show the position of the 56 stars of the Cetus-Palca spectroscopic sample with common orbital behaviour. The blue rectangle highlights the position of the Cetus-Palca stream visible in the lower panel. The stream is decomposed into seven boxes of equal area used to measure the spatial variation of the CMDs in Fig.~\ref{fig:CMDEDR3}. The cyan polygon highlights the position of the potential globular cluster stream that we find in this work. The two purple dashed rectangles highlight the region located outside the Cetus-Palca stream used to show the MW foreground and background CMD. The red dashed ellipses show the positions of two additional structures visible in the lower panel. The light purple band shows the track of the Sgr stream. Finally, the black and white 2D histogram in the background shows the number of good observations along the scan direction of {\it Gaia} ({\sc astrometric\_n\_good\_obs\_al}).} 
\label{fig:posEDR3}
\end{figure*}

\begin{figure*}
\centering
  \includegraphics[angle=0, clip,width=17.5cm]{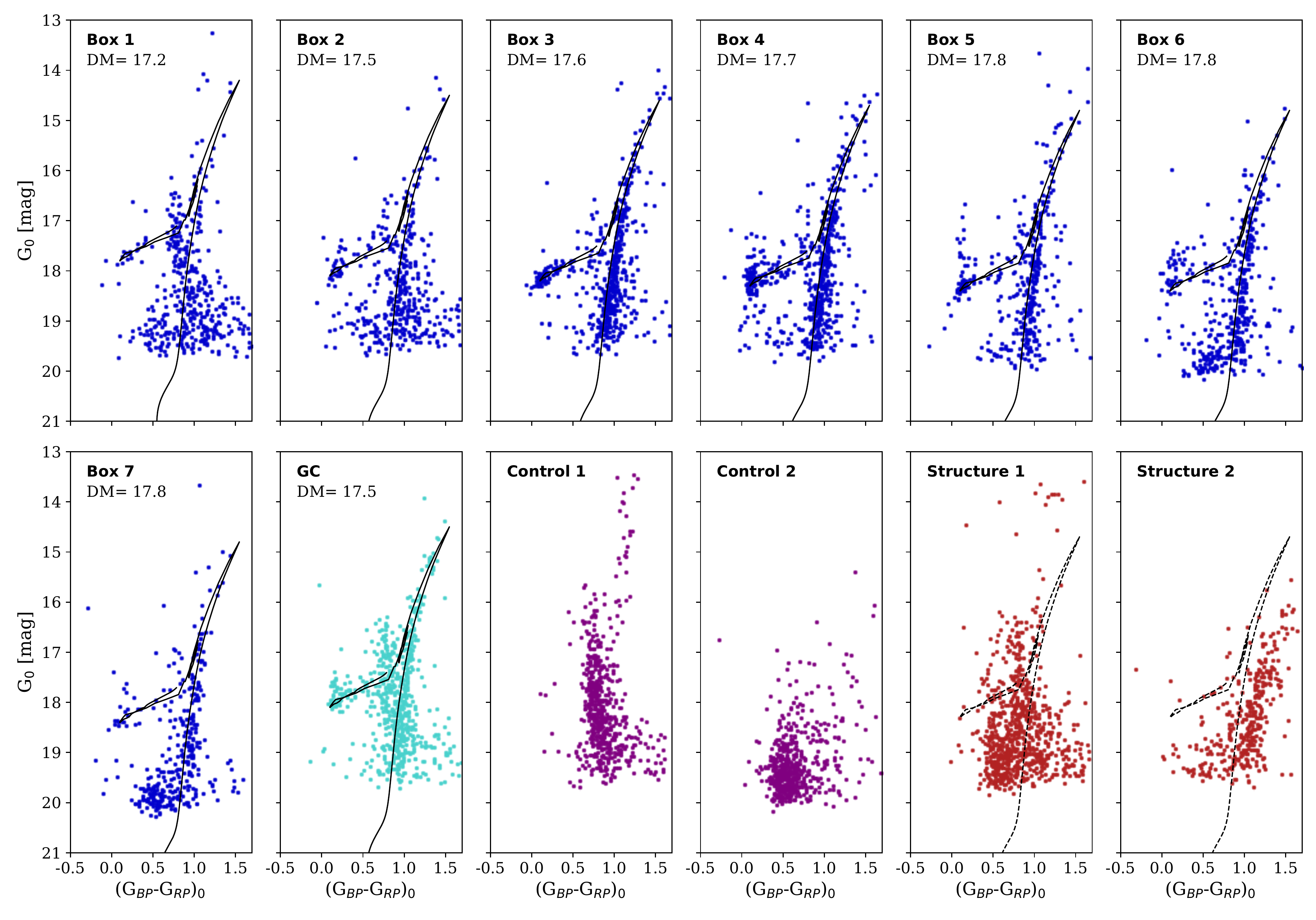}
   \caption{Colour--magnitude diagrams of the different spatial regions shown in Fig.~\ref{fig:posEDR3} in which the black line shows a MIST fiducial isochrone for a 14 Gyr-old population with a metallicity of [Fe/H]=-1.93, shifted by the DM indicated in the legends. For structures 1 and 2, the DM is 17.7. The colours of the symbols reflects the colour coding of the structures highlighted in Fig.~\ref{fig:posEDR3}.}
\label{fig:CMDEDR3}
\end{figure*}

\begin{figure}
\centering
  \includegraphics[angle=0, clip,width=8.0cm]{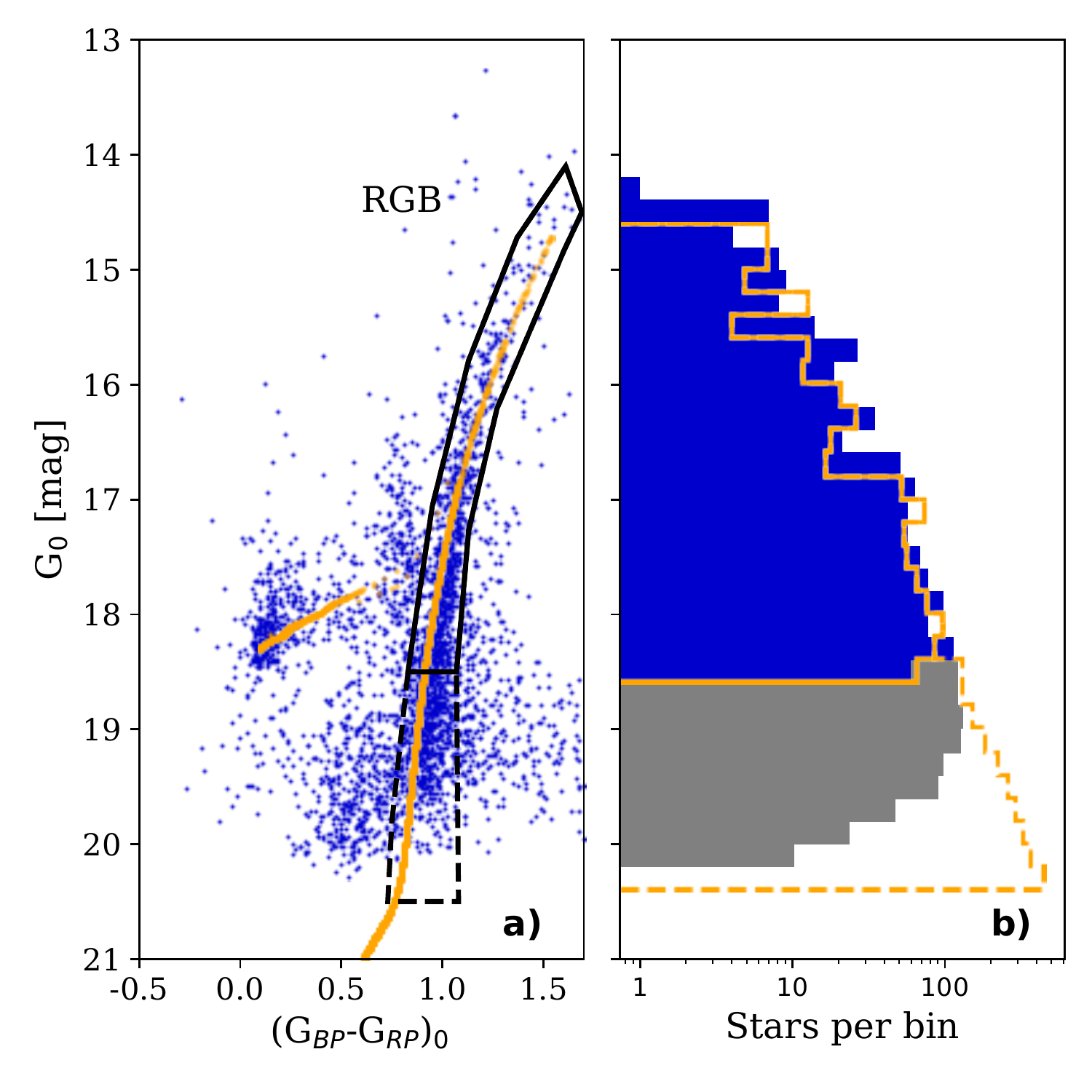}
   \caption{Panel {\it (a):} CMD of the stars located in the Cetus-Palca selection box (blue rectangle in Fig.~\ref{fig:posEDR3}) overlaid with a 14 Gyr isochrone with a metallicity of [Fe/H]=-1.93 (in orange). The black plain polygon shows the location of the RGB stars with the region where {\it Gaia} is complete, while the region where {\it Gaia} is not complete is shown by the dashed line polygon. Panel {\it (b):} Luminosity function of the RGB used to estimate the stellar mass of the Cetus-Palca stream (in blue), with the area affected by the drop in completeness (in grey). The orange histogram shows the luminosity function of a $1.5 \times 10^6$ M$_\odot$, [Fe/H]=-1.93, 14 Gyr-old RGB population following a Kroupa mass function.} 
\label{fig:LumFunc}
\end{figure}

\subsection{Tracing the Cetus-Palca stream with different stellar populations} \label{sec:mass}

As noted in the previous section, BHB stars are perfect tracers of the position and the distance gradient along the Cetus-Palca stream. However, these stars only represent a small fraction of the total number of stars of either a dwarf galaxy or a globular cluster, which renders them unusable for measuring the stellar mass of an object like Cetus-Palca. Therefore, to further explore the properties of the Cetus-Palca stream, we expand the analysis to other stellar populations. For this analysis, we used the information gained in the previous section to select potential members of Cetus-Palca.

 The initial selection of stars is similar to that done for the BHBs, except that the colour selection criteria is broader, with $-0.5 \leq ($G$_{BP}$-G$_{RP})_0 \leq 1.7$. Moreover, we restricted the analysis to $-75 \degr < \phi_1< 75 \degr$ and to stars with $\varpi>-0.4$ mas. As we found that Cetus-Palca is always located at distances farther than 20 kpc, we imposed that all stars should have $\varpi<1/5$ mas (rather than $\varpi -3 \delta \varpi < 1/5$ mas for the BHBs), again in order to remove stars that are very likely from the disc, while keeping as many stars of the halo as possible. Finally, we restrict the analysis to the proper motion plane $-3$  mas.yr$^{-1} \leq \mu_{\phi,1} \leq 2$ mas.yr$^{-1}$ and $-2$ mas.yr$^{-1} \leq \mu_{\phi,2} \leq 3$ mas.yr$^{-1}$.

Our adopted approach to select stars that are  likely members of the Cetus-Palca stream is largely inspired by the methods used to statistically separate the stars of a dwarf galaxy from the background and foreground contamination \citep[e.g.][]{martin_2013,longeard_2018,pace_2019,mcconnachie_2020,mcconnachie_2020a,battaglia_2021}. The general premise of those methods is that any star in {\it Gaia}  is either a member of a stellar structure (e.g. a dwarf galaxy or a stellar stream) or of the Milky Way foreground or background. Therefore, the unmarginalised likelihood $\mathcal{L}$ of a given star can be defined as\begin{equation}
    \mathcal{L}=f_{sub} \  \mathcal{L}_{sub} + (1-f_{sub}) \  \mathcal{L}_{MW} \ ,
\end{equation}
where $\mathcal{L}_{sub}$ and $\mathcal{L}_{MW}$ are respectively the likelihoods of the stellar substructure and of the MW foreground and background, and $f_{sub}$ is the fraction of stars in the stellar structure.

$\mathcal{L}_{sub}$ and $\mathcal{L}_{MW}$ can be decomposed as spatial ($\mathcal{L}_s$) and kinematic ($\mathcal{L}_{PM}$) components. However, contrary to dwarf galaxies that cover a small area of the sky and for which there is no spatial variation of the measured proper motions (at first order and with the current precision of the instrument), this is not the case for an elongated stellar stream such as Cetus-Palca. Similarly, the proper motions of the MW stars vary strongly with position on the sky. Therefore, we decoupled the kinematic component in two terms, where the first one $\mathcal{L}_{PM1}$ is the likelihood of the proper motion at a given position along the first axis, and $\mathcal{L}_{PM2}$ is the likelihood of the proper motion at a given position along a second perpendicular axis. As a result, we obtain $\mathcal{L}_{sub}$ and $\mathcal{L}_{MW}$  expressed as:
\begin{equation}
\mathcal{L}_{x}= \mathcal{L}_{s,x} \ \mathcal{L}_{PM1,x} \ \mathcal{L}_{PM2,x} \ ,
\end{equation}
where the index $x$ refers either to the MW or to the satellite. 
Contrary to the works mentioned above, we did not use colour--magnitude information to compute the probability of being a member of Cetus-Palca because we want to use this as an independent validation check of our method, but also because we use the luminosity function to measure the stellar mass of Cetus-Palca.

For this analysis, we work with the equatorial coordinates both for position and proper motions (R.A., Dec., $\mu_\alpha^*$ and $\mu_\delta$) because we can assume, in a first approximation, that the variation of $\mu_\alpha^*$ along the right ascension axis is decoupled from the variation of $\mu_\delta$ along the declination axis. This would not be the case if we worked in the Cetus-Palca coordinates, because we would have to introduce a correlation matrix for the proper motion perpendicular to the plane of the stream ($\mu_{\phi 2}$) to include its variation along $\phi_1$ and along $\phi_2$.

For the stream likelihood ($\mathcal{L}_{sub}$), the positional proper motion likelihoods in the right ascension and declination planes $\mathcal{L}_{PM1,sat}$ and $\mathcal{L}_{PM2,sat}$ are computed from the BHBs identified as being part of the Cetus-Palca (see Section~\ref{sec:BHB_cetus}). For both planes, we modelled the variation of the proper motion with the position by a third-order polynomial fitted on the distribution of the Cetus-Palca BHBs. In this way, at every position, the distribution of proper motion is modelled by a Gaussian function with a width equal to the residual BHB distribution with respect to the polynomial fit ($\sigma=0.197$ mas.yr$^{-1}$ for the right ascension plane and $\sigma=0.143$ mas.yr$^{-1}$ for the declination plane). As it has been found in previous works that the morphology of the Cetus-Palca stream is rather complex \citep{newberg_2009,yam_2013,yuan_2019}, and because the foreground and background MW contamination is largely dominating the signal, we assumed a uniform distribution  for the spatial likelihood for the stream. For the MW foreground and background distribution, because it largely dominates the signal, we constructed the spatial and the two kinematics likelihood components directly from the distribution of the selected {\it Gaia} stars, and smoothed them over $1\degr$ spatially and over 0.05 mas.yr$^{-1}$ in proper motion space.

The fraction of stars that are part of Cetus-Palca ($f_{sub}$) was found by exploring the parameter space of the posterior distribution with {\sc EMCEE} \citep{foreman-mackey_2013}. The posterior distribution of the data ($D$), given a model with a fraction of stars in Cetus-Palca $f_{sub}$, is defined such as $P(D|f_{sub}) \propto \mathcal{L} \times P(f_{sub})$, where $P(f_{sub})$ is a uniform flat prior on $f_{sub}$ between 0 and 1. Doing so, we found a fraction of stars in Cetus-Palca equal to $f_{sub}=0.00837 \pm{0.00016}$.

Once the value of $f_{sub}$ is found, it is possible to compute the probability of belonging to Cetus-Palca ($P_{Cetus}$) for each star, as
\begin{equation}
P_{Cetus} = \frac{f_{sub} \ \mathcal{L}_{sub}} {f_{sub} \  \mathcal{L}_{sub} + (1-f_{sub}) \  \mathcal{L}_{MW}}\ .  
\end{equation}

The spatial distribution of the 17 990 stars with $P_{Cetus}\geq 0.2$ is shown in the lower panel of Fig.~\ref{fig:posEDR3}. The selection still contains contamination from the MW disk, the LMC, and the Sgr stream, but we can clearly see the extent of the Cetus-Palca stream in the bottom panel, whose position is highlighted by the blue rectangle in the upper panel. We arbitrarily divided the stream into seven boxes of equal area, for which we present the respective CMDs in Fig.~\ref{fig:CMDEDR3}. The RGB of the Cetus-Palca stream is clearly visible in each of these boxes, and has a very different signature from the background, as shown by the CMDs of the two control areas (purple rectangle in Fig.~\ref{fig:posEDR3}). The CMDs of boxes 3 and 4 are by far the most populated. These two boxes cover the region where we believe the centre of the progenitor to be (Section~\ref{sec:BHB_cetus}). For each box, we manually adjusted the  DM of a 14 Gyr MIST isochrone with [Fe/H]=-1.93 \citep{dotter_2016,choi_2016} to the CMD. The age of the isochrone was chosen because it gives the best fit to the colour extent of the BHBs, and the metallicity corresponds to that which we measured with the spectroscopic sample (Section~\ref{phase-space}). The variation in distance measured in this way is highly consistent with that measured from the BHBs. 

The co-added CMD of the seven boxes where we identified Cetus-Palca is shown in panel (a) of Fig.~\ref{fig:LumFunc}, where an MIST isochrone with a distance modulus of 17.7 mag is overlaid. To take into account the variation in distance along the Cetus-Palca stream, before co-addition, the various CMDs were offset by the difference between the DM indicated in each panel of Figure~\ref{fig:CMDEDR3} and the mean distance modulus of Cetus-Palca of 17.7 mag. The RGB and BHB are clearly visible, while the contamination from MW stars is relatively low, because the disc MS is barely detectable. In this figure, the black presents the RGB selection box whose luminosity function is shown in panel (b). We compared the luminosity function of the observed RGB to a fiducial RGB population with a \citet{kroupa_2001} mass distribution to adjust the total luminous mass of Cetus-Palca. For this, we only used the stars brighter than G$_0=18.5$, because the completeness drops rapidly at fainter magnitudes. This corresponds to the threshold magnitude up to which {\it Gaia} can be considered as complete in this region of the sky \citep{boubert_2020}. 
We performed 100 random realisations to take into account the photometric uncertainties of each star. In doing so, we find that the resulting stellar mass ranges between $1.3 - 1.6 \times 10^6$ M$_\odot$, with a median of $1.4 \times 10^6$ M$_\odot$, consistent with the upper mass range found by \citet{yuan_2019}. If we were to consider stars with $P_{Cetus}\geq 0.1$, we would obtain a very similar result, with a stellar mass between $1.3 - 1.7 \times 10^6$ M$_\odot$. This is in the same order of magnitude as `classical' dwarf spheroidal galaxies like Sextans or Sculptor, which have stellar masses of $0.4 \times 10^6$ and $7.8 \times 10^6$ M$_\odot$, respectively \citep{irwin_1995,lokas_2009,deboer_2012}. According to the stellar mass--metallicity relation for Local Group dwarf galaxies from \citet{kirby_2013}, such a massive galaxy should have a metallicity of [Fe/H]=$-1.63\pm{0.16}$. This is consistent with the metallicity that we measured for the Cetus-Palca stream in Section~\ref{Sec_phase_space}, though it is towards the lower limit of the observed stellar mass--metallicity relation by \citet{kirby_2013}. However, this relation has large uncertainties in the metal-poor regime, and a galaxy with a similar stellar mass, namely Andromeda~III, is also found to be more metal poor than predicted by this relation, with [Fe/H]=$-1.84\pm{0.05}$ \citep{kirby_2013}. 

\subsection{Detection of a potential stream of stream}

In addition to the Cetus-Palca stream, another, fainter stream-like structure parallel to the Cetus-Palca stream is visible in Fig.~\ref{fig:posEDR3} located around $\phi_2 \simeq 10\degr$  (highlighted by the cyan polygon in the upper panel). This structure is also visible in the distribution of the BHBs we identified as members of Cetus-Palca (black circles in the upper panel of Fig.~\ref{fig:posEDR3}). The position of this structure does not coincide with an area observed more than others by {\it Gaia} and where the completeness could be higher, as illustrated by the number of good observations in the lateral-scan direction of the {\it Gaia} satellite, {\sc astrometric\_n\_good\_obs\_al}, (2D histogram in the upper panel of Fig.~\ref{fig:posEDR3}). This structure seems to be spatially decoupled from the position of the Cetus-Palca stream and the CMD of the stars in that region (whose area is equal to that of an individual box along Cetus-Palca) is more populated than in any of the neighbouring boxes. 

The small width of this stream ($\approx 300$ pc) could indicate that it might be formed by the tidal debris of a disrupted globular cluster that was orbiting around the progenitor of Cetus-Palca. The number of stars with spectroscopic measurements that fall on this feature and share a common orbital behaviour to Cetus-Palca is too small (7 stars) to establish its nature with certainty. The low scatter in metallicity (0.08 dex around a value  [Fe/H]$=-1.93$dex) would support the globular cluster hypothesis. However, the distance gradient along this feature and the kinematics of its stars are very similar to those of the rest of the Cetus-Palca spectroscopic sample. Only the line-of-sight velocity might be different from those of the other stars located along the main Cetus-Palca stream but with so few measurements, this is not a significant result. Therefore, it is not possible exclude the possibility that this second structure is actually another appendix of the Cetus-Palca stream itself. Such a structure could result, for example, from the non-linear perturbations caused by the LMC \citep[see][for discussions on the influence of the LMC on stellar streams]{erkal_2019,shipp_2019,vasiliev_2021}.

The spatial location and the distance of $\simeq 31$ kpc of this secondary stream are consistent with the already known Triangulum/Pisces stream \citep[Tri/Psc;][]{bonaca_2012,martin_2013a,martin_2014}. However, the secondary stream found in our study is significantly more extended than previously found ($\simeq 50\degr$ against $12\degr$ in \citealt{bonaca_2012} and $13 \degr$ in \citealt{martin_2014}). The presence of this secondary stream on Fig.~\ref{fig:posEDR3} along the Cetus-Palca stream suggests that the Tri/Psc and the Cetus-Palca streams share a common origin, as recently proposed by \citet{bonaca_2021} and \citet{yuan_2021}.

To confirm (or disprove) the hypothesis that the Tri/Psc stream formed by the disruption of a globular cluster that was initially around the progenitor of Cetus-Palca, spectroscopic follow-up is required to confirm that those stars do not present a scatter in metallicity, have a similar kinematic behaviour, and present the CNONaAl abundances anti-correlation typical of old globular clusters \citep{carretta_2010,bastian_2018}. Should it be confirmed that this second structure is a second stream, to the best of our knowledge, this would be the first detected tidal stream of an object orbiting around a dwarf galaxy that itself is currently being disrupted. Such a discovery would be particularly useful in constraining the initial profile of the dark matter halo of the Cetus-Palca dwarf galaxy \citep{malhan_2021}.

Two other stream-like structures are visible in Fig.~\ref{fig:posEDR3}, illustrated by the red ellipses in the upper panel. However, regarding the CMD of Structure 1 in Fig.~\ref{fig:CMDEDR3} we can see that it is a composition of the CMDs of the two control regions, which indicates that it is a structure of the disc, likely an artefact due to the high number of stars in that region close to the plane of the disc. For Structure 2, a RGB more distant than that of Cetus-Palca is visible, which actually corresponds to the Sgr stream, whose track is shown by the light purple band in Fig.~\ref{fig:posEDR3}. 
%
%\begin{figure}
%\centering
%  \includegraphics[angle=0, clip,width=8.0cm]{fig15.pdf}
%   \caption{{\it Upper panel:} Line-of-sight velocity trend along the Cetus-Palca %stream of the 56 stars of the spectroscopic sample with common orbital behaviour. The %stars located in the candidate globular cluster stream are represented by the cyan %triangles, while the 49 others are represented by the orange stars.{\it Lower panel:} %Heliocentric distance trend along the Cetus-Palca strean for the same stars as in the %upper panel.} 
%\label{fig:VlosTrend}
%\end{figure}

\section{Summary and Conclusions} \label{sec:conclusion}
We present a new method to derive  accurate spectro-photometric distances with a high relative precision  based heavily on supervised machine learning. Specifically, the method uses the $griz$ photometric bands from the Pan-STARRS 3$\pi$ survey, the $G$-band from {\it Gaia} EDR3, and the effective temperature, surface gravity, and metallicity derived from a given spectroscopic survey. This method will be used to derive distances for the stars observed by the Galactic Archaeology  WEAVE surveys \citep{dalton_2012} through the {\sc SPdist} {\it Contributed Data Product}. Meanwhile, we applied the technique to SDSS and SEGUE data and  derive the distance of approximately 300 000 stars in different evolutionary phases; the catalogue is available on the Vizier service of the CDS. In its application to SDSS and SEGUE data, the relative precision on the spectro-photometric distance is of 13\%.  

With the 6D sample obtained using these spectro-photometric  distances, SEGUE line-of-sight velocities, and {\it Gaia} EDR3 proper motions, we searched the integrals-of-motion plane and found a structure corresponding to the Cetus-Palca stellar stream \citep{newberg_2009}, hidden below the location of Sagittarius in the azimuthal and vertical actions (and energy and vertical angular momentum). However, this stellar structure is clearly distinct from Sagittarius in its proper motion and radial velocity in the RA range $10-35 \degr$. 

The good distance precision allowed us to integrate the orbits of these stars and to find the plane of the Cetus-Palca stream, which we then used to perform a detailed analysis, expanding the search for the Cetus-Palca stream to the whole sky using the 5D {\it Gaia} EDR3 sample. We measured a total extent  of the stream of $\simeq 100 \degr$ on the sky over $0$ to $100 \degr$ in R.A. and over  $-80$ to $65 \degr$ in DEC., overlapping  with the Palca over-density detected by \citet{shipp_2018} in the Dark Energy Survey, as predicted by \citet{chang_2020}. However, we did not find a northern counterpart, as suggested in the work of \citet{yuan_2019}, possibly because of its diffuseness. The detected stream covers a heliocentric distance range of between 25 and 40 kpc, is metal poor, and presents  a small scatter in metallicity ([Fe/H]$=-1.93 \pm{0.21}$) typical of a dwarf galaxy and consistent with previous measurements.

Our analysis of the luminosity function leads us to estimate the stellar mass of the Cetus-Palca progenitor at $1.5 \times 10^{6}$ M$_\odot$. This stellar mass is consistent with expectation from the stellar mass--metallicity relation of Local Group dwarf galaxies.
 
 We also report the discovery of a second structure almost parallel to the Cetus-Palca stream, covering $\sim 50\degr$ of the sky and located at a similar heliocentric distance to Cetus-Palca. This second structure appears to be a longer segment of the already known Tri/Psc stream. The finding of this stream along the Cetus-Palca stream suggests that it is emanating from a globular cluster that was initially orbiting around the Cetus-Palca progenitor and is now completely disrupted, as recently proposed by \citet{bonaca_2021} and \citet{yuan_2021}.

\begin{acknowledgements}
The authors thank Rodrigo Ibata and Eduardo Balbinot for useful discussions, as well as the anonymous referee for comments that increased the clarity of the publication.

GT acknowledges support from the Agencia Estatal de Investigaci\'on (AEI) of the Ministerio de Ciencia e Innovaci\'on (MCINN) under grant FJC2018-037323-I. The authors acknowledge financial support through the
grant (AEI/FEDER, UE) AYA2017-89076-P, as well as by the Ministerio de
Ciencia, Innovación y Universidades (MCIU), through the State Budget and by
the Consejería de Economía, Industria, Comercio y Conocimiento of the Canary
Islands Autonomous Community, through the Regional Budget. 

\end{acknowledgements}

\bibliographystyle{aa}
\bibliography{biblio}

\begin{appendix}

\section{Spectro-photometric distances with a different training set} \label{app_A}

\begin{figure*}
\centering
  \includegraphics[angle=0, clip, viewport= 0 0 1025 640,width=17.5cm]{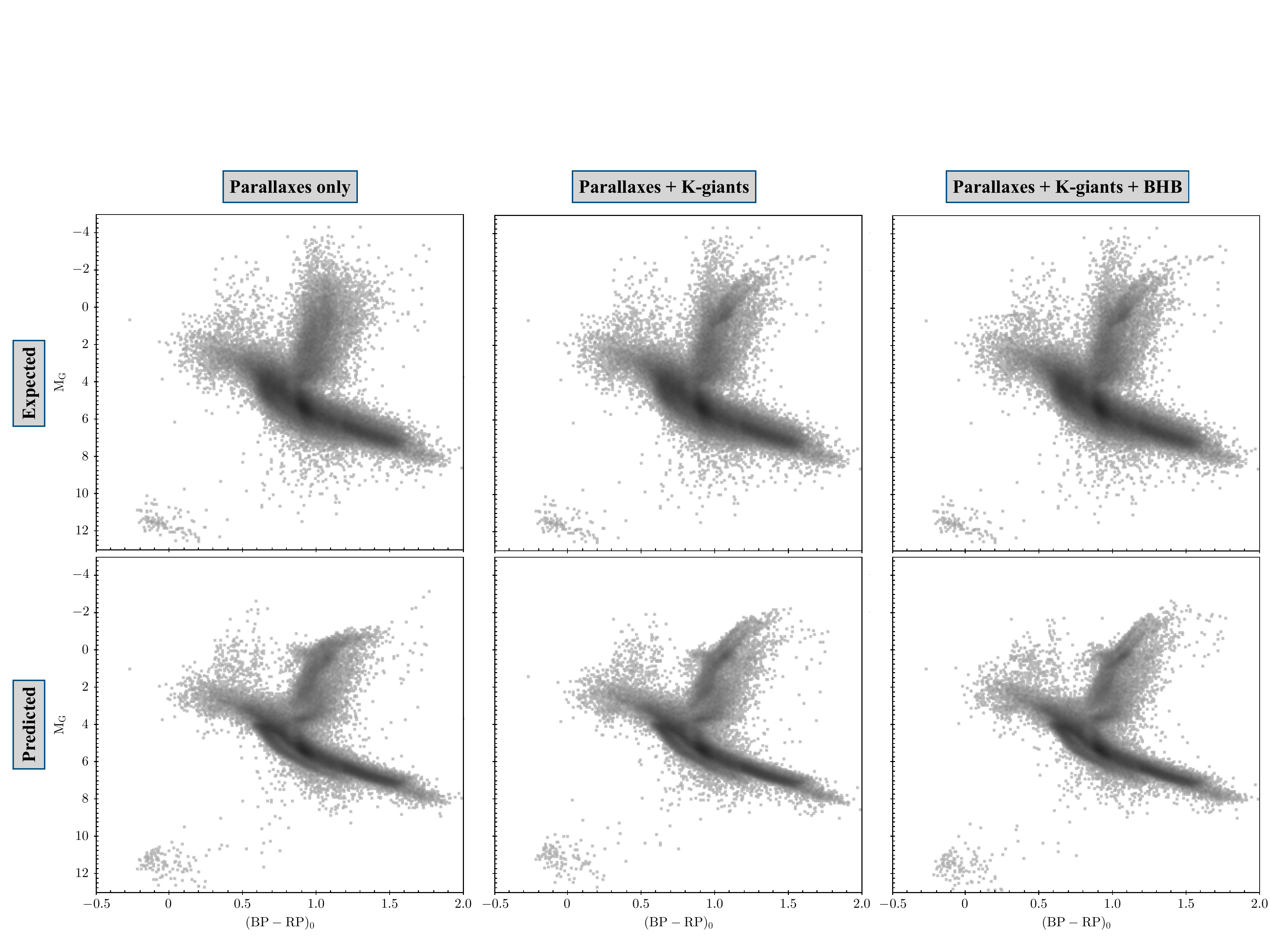}
   \caption{Expected and predicted colour-magnitude diagram for the different training set we explored. Upper panels show the expected (BP-RP)$_0$ -- M$_G$ CMDs for a training sample composed only of stars that respect the parallax criteria presented in Section~\ref{sec:method} (on the left), with the inclusion of K-giants (in the middle), and with the inclusion of both K-giants and BHBs (on the right). We only show the stars in common between the different samples, because the number of stars used to train the ANN are different from one training sample
to another because of the different constraints used to create them. The lower panels show the corresponding CMDs with the absolute magnitude M$_G$ predicted by the ANN. We can clearly see the improvement made by the inclusion of the different catalogues on the global shape of the CMD, especially from the K-giants.} 
\label{fig:annex}
\end{figure*}

Here, we present the improvement on the distance measurement made the by the addition of the K-giant catalogue of \citet{xue_2014} and of the BHB catalogue from \citet{xue_2008} to the training sample of the ANN used to derive spectro-photometric distances, which motivated our choices in the construction of the final training sample used by the ANN presented in Section~\ref{sec:spdist}.

Figure~\ref{fig:annex}  presents the expected and predicted CMDs for the different training samples made with the addition of K-giants and BHBs to stars selected from their parallax. The general accuracy of the ANN is the same for the different training samples because they are composed mainly of MS stars (>80\%, but the actual number differ slightly between the training samples), and these are the same between the different training samples. However, the global shape of the CMDs is different, especially for the top part of the RGB (M$_G \lessapprox -1$). Indeed, without the inclusion of the K-giants to the training sample, the predicted absolute magnitudes tend to a plateau around M$_G= -1$, and the RGB is globally less well-defined. This is because in the training sample based only on the parallaxes, the stars on the RGB have a low relative precision on their parallax. Nevertheless, it is interesting to see that for all cases, the ANN improves the shape of the predicted CMD compared to the expected one, with a clear separation between giants and subgiants for example (around $0.8 \leq$(BP-RP)$_0\leq 1.2$ and $2\leq$ M$_G \leq 4$).

The inclusion of the BHBs in the training sample has a small effect in that it leads to a better definition of the horizontal branch. However, its effect is far from marginal, since without its inclusion the absolute magnitude predicted for the BHB stars is on average $\simeq 0.2$ magnitude fainter than measured with the distance calibration of \citet{deason_2011}, leading to systematic overestimation of the distance (or accuracy error) by $\simeq 10\%$.

\section{Precision of the spectro-photometric distances as function of T$_{eff}$, $\log(g),$ and [Fe/H]} \label{app_B}

\begin{figure*}
\centering
  \includegraphics[angle=0, clip,width=17.5cm]{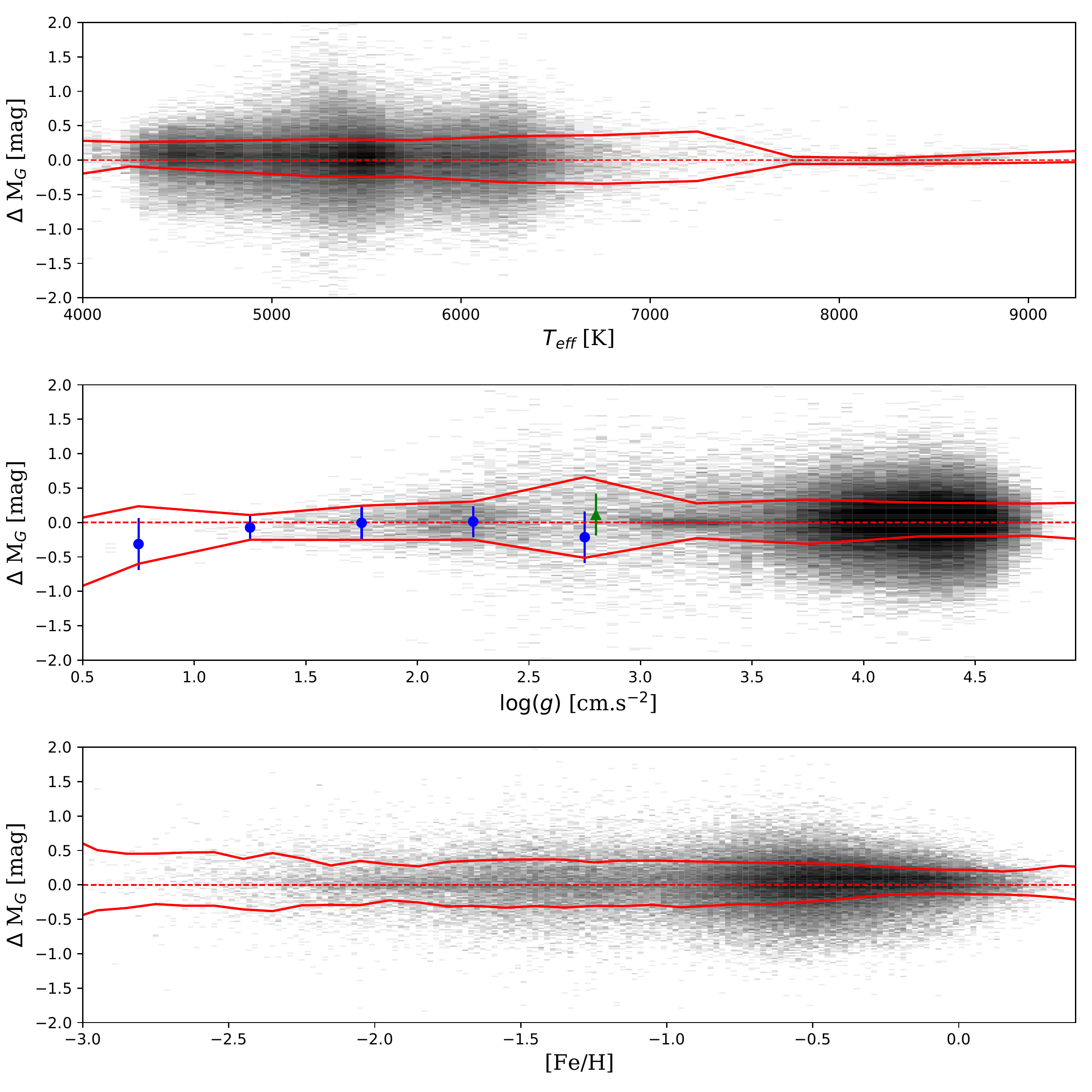}
   \caption{Variation in the residual between the `true' and predicted absolute magnitudes with the effective temperature (upper panel), the surface gravity (middle panel), and the metallicity (lower panel). The dotted line shows cases where the predicted and the `true' absolute magnitude are equal. The red continuous line shows the variation in this residual as a function of these parameters. In the middle panel, the blue dots show the evolution of the residual for the K-giant stars of the catalogue of \citet{xue_2014}, and the green triangle shows it for the 229 stars between $2.5<\log(g)<3.0$ dex and with relative precision on their parallax  of $\varpi/\delta \varpi \geq 10$.} 
\label{fig:annexb}
\end{figure*}

The mean relative precision on the distance predicted by the ANN presented in Section~\ref{sec:method} is 13\%, or 0.27 mag, but the actual precision varies for stars with different effective temperature, surface gravity, and metallicity. To analyse the precision of the spectro-photometric distances over these three parameters, we selected stars of the training sample located in different regions of the parameter space, and we fitted a Gaussian function to their $\Delta M_G = $M$_{true}$ - M$_{predicted}$ histogram. The variation of the precision of the residual absolute magnitude with the effective temperature, surface gravity, and metallicity is presented in Figure~\ref{fig:annexb}.

\paragraph{Precision as a function of the effective temperature}
The variation of the residual between the `true' and predicted absolute magnitude as a function of the effective temperature is presented in the upper panel of Figure~\ref{fig:annexb}, where the bins in effective temperature have a width of 500 K. One can see that between $4000<$ T$_{eff}<7500$ K, the residual is relatively flat with a typical scatter of 0.26 mag, slightly rising to a scatter of 0.36 mag at T$_{eff}=7250$ K due to the low number of stars in that range, making a good fit to the distance in that range more difficult. Beyond 7500 K, the precision increases drastically, with the scatter between the `true' and the predicted absolute magnitude going down to  0.05 mag. This is because in that range, more than two-thirds of the stars of the training sample are BHBs that have a low intrinsic scatter \citep[$<0.1$ mag][]{deason_2011}, making it easier to obtain a precise predicted absolute magnitude for these stars.

\paragraph{Precision as function of the surface gravity}
The middle panel of Figure~\ref{fig:annexb} shows the evolution of the residual absolute magnitude with the surface gravity, where the bins in surface gravity have a width of $\log(g)=0.5$ dex. The residual is relatively constant over the range of surface gravity covered by SEGUE, except toward $\log(g)=0.5$ dex where the predicted magnitudes are overestimated by 0.18 mag and the scatter of the residual goes to 0.42 mag. This is due to the very low number of stars in that range and to the poor determination of the surface gravity by the SSPP in that range, as illustrate the numerous clumps of stars with exactly the surface gravity values in that surface gravity range. It should be noted here that the trend for the K-giants, shown by the blue dots, is similar to the trend of the global training set for $\log(g)>2.5$ dex because they constitute the majority of the training sample in that range. Between $2.5<\log(g)<3.0$ dex,  the difference between the `true' and the predicted absolute magnitude is also abnormally high, with a scatter of 0.58 mag. In that range of surface gravity, the large majority of the stars of the training sample (87\%) have been selected based on their absolute parallax precision ($\delta \varpi < 0.07$ mas), and are not part of the K-giant catalogue of \citet{xue_2014} or of the BHB catalogue of \citet{xue_2008}. For this reason, the `true' absolute magnitudes of the stars of the training sample are not well-defined in that surface gravity range, with an average uncertainty on the `true' absolute magnitude of $\delta M_G=0.94$ mag, rather than $\delta M_G=0.26$ mag for the global training sample. This low precision of the `true' absolute magnitudes alone explains the increase in the scatter of the residual. However, the precision on the predicted absolute magnitude in that range of surface gravity is likely similar to that obtained for the rest of the stars because we found a scatter of the residual of 0.30 mag for the 229 (over $3,715$) stars with $\varpi/\delta \varpi \geq 10$ located in that region, but with predicted absolute magnitudes systematically underestimated by 0.1 mag, as showed by the green triangle in Figure~\ref{fig:annexb}. In comparison, for the 355 K-giants located in that region, the predicted absolute magnitudes are systematically overestimated by 0.2 mag, with a scatter on the residual of 0.37 mag. However, this comparison has to be taken with caution because the K-giants in that region have a typical uncertainty on their `true' absolute magnitude of $\delta$ M$_G=0.42$ mag, which is higher than for the rest of the K-giants (M$_G=0.32$ mag). Regarding the results of this analysis, we cannot definitively conclude that the precision on the absolute magnitude (distance) between $2.5<\log(g)<3.0$ dex is similar to that obtained for other surface gravity ranges, and the distances estimated for the stars in that range also have to be taken with caution. 

\paragraph{Precision as function of the metallicity}
The variation residual absolute magnitude between the `true' and the predicted one as a function of the metallicity is presented in the lower panel of Figure~\ref{fig:annexb}, where the bins in metallicity have a width of [Fe/H]=0.1. The scatter is constant (typically of 0.28 mag) over almost the full range of metallicities, with a small increase towards the low metallicity edges caused by the low number of stars in that region and the imprecision of the metallicities predicted by the SSPP for extremely metal-poor stars \citep[see][]{fernandez-alvar_2016,starkenburg_2017}. Around [Fe/H]$\simeq -0.1$, the scatter on the residual is of 0.17 mag, which is lower than for other metallicities because in that region the stars of the training sample have a typical uncertainty on the `true' absolute magnitude of $\delta$ M$_G=0.15$ mag, which is lower than for the rest of the stars of the training sample, which have $\delta$ M$_G=0.26$ mag.

\end{appendix}
\end{document}